\documentclass[aps,prb,twocolumn,showpacs,superscriptaddress,groupedaddress,reprint]{revtex4-1}  % for review and submission
\usepackage{graphicx}  % needed for figures
\usepackage{dcolumn}   % needed for some tables
\usepackage{bm}        % for math
\usepackage{amsmath,tabularx}   % for math
\usepackage{subcaption}
\usepackage{float}

\begin{document}
% the following line is for submission, including submission to the arXiv!!
%\hspace{5.2in} \mbox{SLAC-Pub-XX/xxx-x}

%\title{An efficient source interpolation method for the numerical solution of steady state self consistent beam-wave interactions}
\title{A classical field theory formulation for the numerical solution of time harmonic electromagnetic fields}
\author{A.~Gold}
\email{vrielink@stanford.edu}
\affiliation{Stanford University, Stanford, California 94305, USA}
\author{ S.~Tantawi}
\affiliation{SLAC National Accelerator Laboratory, Menlo Park, California 94025, USA}
\date{\today}

\begin{abstract}
Finite element representations of Maxwell's equations pose unusual challenges inherent to the variational representation of the ``curl-curl'' equation for the fields. We present a variational formulation based on classical field theory. Borrowing from QED, we modify the Lagrangian by adding an implicit gauge-fixing term. Our formulation, in the language of differential geometry, shows that conventional edge elements should be replaced by the simpler nodal elements for time-harmonic problems. We demonstrate how this formulation, adhering to the deeper underlying symmetries of the four-dimensional covariant field description, provides a highly general, robust numerical framework. 
\end{abstract}
\pacs{}
\maketitle

\section{Introduction}
\label{sec:Introduction}
As physicists and engineers seek to model increasingly complex electromagnetic systems, from radio-frequency power sources to integrated photonics, the need for efficient and robust full-wave, first-principles numerical field solvers is growing. Finite element (FE) methods, which solve partial differential equations over a discretized problem domain (often a spatial mesh), are a natural solution enjoying widespread use in disciplines from fluid dynamics to structural mechanics. 

In electromagnetic problems, the variational formulation driving the FE method presents unique challenges which impede the computational efficiency and accuracy of existing solvers, however. For time harmonic problems where the fields oscillate at angular frequency $\omega$, $\vec{\mathcal{E}}(\vec{r},t) = \mathrm{Re}[\vec{E}(\vec{r}) e^{i \omega t}]$, this variational expression is given by eq.~\ref{eq:CC}. Here, $\mu_r$ and $\epsilon_r$ are the relative permeability and permitivitty, $k_0$ is the wave number in free space, $Z_0$ is the intrinsic impedance of free space and $\vec{J}$ is the current density. For a more detailed treatment see, for example, ref. \onlinecite{FEMtextbook}.
\begin{equation}
F=\int_{\Omega}  \frac{1}{\mu_r} | \nabla \times \vec{E}|^2 - k_0^2 \epsilon_r |\vec{E_r}|^2 + i k_0 Z_0 (\vec{E}^{*} \cdot \vec{J}-\vec{E} \cdot \vec{J}^{*} )   ~\mathrm{d}V
\label{eq:CC}
\end{equation}

The primary challenge consists in enforcing the divergence constraint associated with Gauss's law, $\nabla \cdot \vec{E} =0$ (in the source-free case) while ensuring adequate freedom in the basis functions used to expand the approximate solution so as to be able to model discontinuities in the fields \cite{ARreview,ICGreview}. Alternatively, when working with the magnetic vector potential, $\vec{A}$, the divergence requirement is necessary to enforce the Coulomb gauge,  $\nabla \cdot \vec{A} = 0$. These two requirements conflict in standard nodal element based finite element (FE) methods, employed successfully in other fields such as fluid mechanics and structural mechanics.

In problems with charge and current density, $\rho$ and $\vec{J}$, there is the added dilemma of how best to satisfy both the Ampere-Maxwell equation and Gauss's law. In particle-in-cell codes, used in plasma and accelerator physics, this is critical as discrete charge conservation is not automatically guaranteed. Correction schemes must be applied to either the field calculation or the source deposition to bound the error in $\nabla \cdot \vec{E}$ and avoid any resulting numerical instabilities. \cite{marder,LangdonGaussLaw,CiarletGaussLaw,GeneralizedFormulationForVlasovMaxwell,EdgeElementCiarlet} For stationary and low-frequency or broadband problems, such as in electro-quasistatics and integrated circuit design, mixed finite-element solutions combined with tree-cotree splitting of the mesh and/or Lagrange multipliers are commonly applied to account for the contributions from $\rho$ and $\vec{J}$ in the static limit separately \cite{LorenzGaugedMixed, hiptmair,GeneralizedFormulationForVlasovMaxwell}. 

Historically, the application of differential geometry in three dimensional (3D) Euclidean space has sucessfully resolved the first issue, providing a theoretical motivation for the use of edge elements in the expansion of the fields. Despite this success, to the authors' knowledge, the full four dimensional (4D) covariant framework has not been investigated within the context of numerical electromagnetism. We demonstrate herein that such an extension is not only more naturally suited to numerical analysis, the field theory Lagrangian providing a variational form directly applicable to the finite element method, but resolves entirely the two significant issues discussed above. 

We propose a formulation for the finite element solution of electromagnetic systems based on the classical field theory Lagrangian with a gauge fixing term adapted from quantum electrodynamics. This is given in the framework of differential geometry by eq.~\ref{eq:dfGFA}, where we adopt the standard terminology given by introductory texts such as ref. \onlinecite{diffGeoTextbook}. As will be introduced in greater detail shortly, the first two terms constitute the classical Lagrangian and the final term, multiplied by the scalar $\frac{1}{2\xi}$, is the gauge fixing addition. 
\begin{equation}
\mathcal{L}=-\frac{1}{2 \mu} \mathrm{d} \mathbf{A} \wedge  \star \mathrm{d}\mathbf{A} + \mathbf{A} \wedge  \mathbf{J} - \frac{1}{2 \xi \mu } \mathrm{d}\star \mathbf{A} \wedge \star \mathrm{d}\star \mathbf{A}  \label{eq:dfGFA}
\end{equation}
Here $\mu$ is the magnetic permeability, $\mathbf{A}=A_{\nu} \mathrm{d}x^{\nu}$ is the 4D differential 1-form, $A^{\nu}=(\frac{\phi}{c},\vec{A})$ is the four-potential comprised of the electrostatic potential $\phi$ and magnetic vector potential $\vec{A}$, and $\mathbf{J}$ is the electric current 3-form. In this paper we use the metric signature ($+---$). Instead of taking the variation of eq.~\ref{eq:CC} to obtain $\vec{E}$, we propose taking the variation of the action $
S[\mathbf{A}]=\int \mathcal{L}$. This formulation fully accounts for the charge density, $\rho$ in addition to the current $\vec{J}$ through $\mathbf{J}$ and facilitates a return to the widely used nodal FE framework.

The paper is organized as follows: section \ref{sec:3DEuclidean} provides the background for this work, including a more in depth discussion of the edge elements and their benefits and challenges. We then move from 3D Euclidean space into 4D Minkowski space, presenting the classical field theory Lagrangian in section \ref{sec:4DMinkowski}. We demonstrate how this formulation is related to existing $\vec{A}-\phi$ approaches obtained by substituting the potentials into eq. \ref{eq:CC} yet differs in a few critical points which ensures a fully 4D formulation and enables the use of nodal elements instead of edge elements. Finally, in section \ref{sec:Implementation} and \ref{sec:Results} we introduce a proof of concept implementation and demonstrate the validity of the formulation, benchmarking it against a state of the art edge element field solver in terms of accuracy, numerical robustness and flexibility. 

\section{Background}
\label{sec:3DEuclidean}
We motivate this idea by considering how the challenges inherent to eq.~\ref{eq:CC} are currently resolved. Enforcing the divergence constraint was initially addressed by adding a regularization term of the form $s (\nabla \cdot \vec{E})^2$ to eq.~\ref{eq:CC}, with limited success\cite{penaltymethod1983,penaltymethod,FEMmaxwellChapter,ETHEigenSolver}. While the regularization term eliminates spurious non-solenoidal modes in the solution spectrum, the regularized formulation fails to converge to the correct solution for problem geometries with sharp or re-entrant corners. The explanation for this failing was only recently understood: when nodal basis functions are used in conjunction with the regularization term, the approximate solution space spanned is overly restrictive on non-convex domains \cite{CDlong}. When singularities in the field exist such as at sharp corners, it can be shown that the missing subspace consists of the gradients of solutions to Laplace's equation on the same domain\cite{CDarticle,AssousSingularDecomp}. Instead of converging to the correct solution with field singularities, the solution obtained will be the projection of the correct fields on the smooth approximate solution space. 

Two approaches exist to resolve this issue. One can either supplement the nodal basis functions with additional singular or non-conforming functions \cite{SingFieldMethod,AssousSingular,bubbleElements}, or relax the regularization term near the singularity \cite{CDweighted, CiarletWeighted, otin2010regularized}. The former requires computing the coupling between the nodal basis and the additional singular functions and is challenging to extend to three dimensions while the latter is a compromise between enforcing the divergence constraint over the problem domain and not completely restricting the subspace spanned by gradients. 

A more robust solution arises by formulating electromagnetism in the language of differential geometry. In 3D Euclidean space, $\vec{E}$ (and $\vec{A}$ in the Coulomb gauge) are both differential 1-forms which should be expanded not on nodes but on edges. This led to the development and widespread adoption of the ``edge elements'' for electromagnetic problems, as developed separately by Whitney \cite{WhitneyBook} and N\'ed\'elec \cite{NedelecElements}. By their construction, only tangential continuity is imposed at the faces between elements, resolving the issue of modeling field discontinuities at interfaces and boundaries. However, while the edge elements are divergence-free locally, the discontinuity in the normal field at element interfaces allows for solutions that are not divergence-free globally. The space spanned by edge elements divides into the desired space of weakly divergence-free fields and its co-domain, the kernel of the curl operator (purely gradient functions in topologically trivial domains) \cite{WhitneyForms, BossavitTextbook}. There are methods to extract the gradient field so as to span only the divergence-free fields, such as the tree-cotree method \cite{AlbaneseRubinacciTCT,TrappEigenvalueTCT,WangTimeDomainTCT,magnesTCToverview}. However, choosing an optimal tree is challenging and poor conditioning of the resulting linear system is a common issue \cite{TicarBiro,GoliasTreeChoice,PreisNodalEdgeTbad,AhagonKameari}.

\section{Covariant 4D Field Theory Formulation}
\label{sec:4DMinkowski}
\subsection{Insight from Differential Geometry}
Compared to the 3D formulation, electromagnetic theory is encoded much more succinctly by differential geometry in 4D Minkowski space, where deeper underlying structure is made explicit, such as gauge and Lorenz invariance. For Lorenzian manifolds, the equations for the electromagnetic field tensor, $\mathbf{F}=\mathrm{d} \mathbf{A}$, are given by eqs.~\ref{eq:fieldDiff1}-\ref{eq:fieldDiff2}. As defined in section \ref{sec:Introduction}, $\mathbf{A}=A_{\nu} \mathrm{d}x^{\nu}$ is the four dimensional (4D) differential 1-form, $A^{\nu}=(\frac{\phi}{c},\vec{A})$ is the four-potential, and $\mathbf{J} = - \rho \mathrm{d}x \wedge \mathrm{d}y \wedge \mathrm{d}z + j_x \mathrm{d}t \wedge \mathrm{d}y \wedge \mathrm{d}z + j_y \mathrm{d}t \wedge \mathrm{d}z \wedge \mathrm{d}x + j_z \mathrm{d}t \wedge \mathrm{d}x \wedge \mathrm{d}y $ is the current 3-form.
\noindent\begin{minipage}{0.4\columnwidth}
\begin{equation}
\mathrm{d} \mathbf{F} = 0 \label{eq:fieldDiff1}
\end{equation} 
    \end{minipage}%
    \begin{minipage}{0.2\columnwidth}\centering     
    \end{minipage}%
    \begin{minipage}{0.4\columnwidth}
\begin{equation}
\mathrm{d} \star \mathbf{F} =  \mathbf{J} \label{eq:fieldDiff2}
\end{equation}
    \end{minipage}\vskip1em

The classical field theory Lagrangian which encodes these equations, written in terms of $\mathbf{A}$, is given by eq.~\ref{eq:fieldLag}.
\begin{equation}
\mathcal{L}=-\frac{1}{2 \mu} \mathrm{d} \mathbf{A} \wedge  \star \mathrm{d}\mathbf{A} + \mathbf{A} \wedge  \mathbf{J} \label{eq:fieldLag}
\end{equation}

In the covariant treatment it is natural to work with $\mathbf{A}$ as opposed to $\mathbf{E}$ and $\mathbf{B}$, the 3D components of the two-form $\mathbf{F}$. In this case, eq.~\ref{eq:fieldDiff1} is automatically satisfied as $\mathrm{d}^2=0$. Additionally, the charge density, $\rho$, which does not enter into the conventional variational expression for the fields given by eq. \ref{eq:CC}, is accounted for in eq. \ref{eq:fieldLag} through $\mathbf{J}$.

Applying the variational formulation given by the action of eq. \ref{eq:fieldLag} in the finite element method, we first consider the appropriate elements over which to expand the solution, $\mathbf{A}$. As in 3D Euclidean space, where there is a duality between 1-forms and edges, in a 4D mesh the 1-form $\mathbf{A}$ should be expanded using edge elements. This is an intriguing idea to pursue for transient numerical analysis, where one could envision using this formulation on a 4D mesh. However, in the case of time harmonic problems, the focus of this paper, we pursue a different course. The time dimension of the mesh is collapsed and edges of the 4D mesh become points in a 3D mesh. As such, \emph {nodal} elements should be employed to expand $\mathbf{A}$ rather than edge elements.

\subsection{Gauge Invariance and Ill-Conditioned Systems}
\label{GIsection}
The linear systems resulting from employing eq.~\ref{eq:fieldLag} in the finite element method are, unfortunately, highly ill-conditioned. This should come as no surprise when considering that unlike the fields, the four potential is not uniquely defined. The Lagrangian is invariant to gauge transformations of the form $\mathbf{A} \rightarrow \mathbf{A} + d\psi$ where $\psi$ is a scalar 0-form. To resolve this issue, we apply a solution used to address a similar challenge in quantum electrodynamics (QED). In the field discretization in QED, the resulting symbolic matrices are singular due to gauge invariance. One approach to overcome this issue is through the addition of a gauge fixing term to the Lagrangian \cite {QFTSchwartz}:
\begin{equation}
\mathcal{L}_\mathrm{GF}= - \frac{1}{2 \mu \xi} \mathrm{d}\star \mathbf{A} \wedge \star \mathrm{d}\star \mathbf{A} \label{eq:dfGF1}
\end{equation}

The resulting action integral for time harmonic problems, integrated over the time dimension already, is given by eq. \ref{eq:THaction}. Here we have expanded $\mathbf{A}$ into the conventional three-plus-one notation ($\vec{A}$ + $\phi$) for ease of comparison to existing formulations and to expose some implementation challenges which will be discussed in section \ref{sec:Implementation}. Note also that as we are now working in the frequency domain, $\phi$ and $\vec{A}$ as given below are complex quantities. The strong problem corresponding to eq.~\ref{eq:THaction}, obtained by taking the variation of this action integral, is derived in the appendix.
\begin{multline}
S(\phi, \vec{A}) = \frac{1}{2} \int \epsilon |\nabla \phi + i \omega \vec{A}|^2 - \frac{1}{\mu} |\nabla \times \vec{A}|^2 \\-\frac{1}{\mu \xi} |\nabla \cdot \vec{A} - \frac{i \omega}{c^2} \phi|^2 - \rho \phi^{*} -\rho^{*} \phi + \vec{A} \cdot \vec{J}^{*} + \vec{A}^{*} \cdot \vec{J} ~\mathrm{d}V
\label{eq:THaction}
\end{multline}

In QED, the gauge fixing term imposes different gauges depending on the value of $\xi$. In our classical context, any $\xi \neq 0$ imposes the Lorenz gauge with residual gauge freedom, $\mathbf{A} \rightarrow \mathbf{A} +\mathrm{d}\psi$ for $\psi$ satisfying the wave equation $\nabla^2 \psi + k_0^2 \psi = 0$. By setting the components of $\mathbf{A}$ or their derivatives explicitly on the boundary, $\psi$ is forced to zero on the boundary, and hence everywhere, and $\mathbf{A}$ will be unique.  
 
Tempting as this may be, doing so in the most straight-forward way (setting $\vec{A}_t$ =0 and $\phi=0$ on the boundary for a perfect electric conductor, or $\vec{A}_n$ =0  and $\nabla_n \phi =0$ for a perfect magnetic conductor) in fact decouples $\phi$ from $\vec{A}$. This reduces the problem to two separate wave equations: eq.~\ref{eq:phiOnly} amenable to nodal elements and eq.~\ref{eq:AOnly} which reduces to the curl-curl equation for $\vec{A}$ and must be solved via edge elements. 
\begin{align}
\nabla^2 \phi + k_0^2 \phi = -\frac{\rho}{\epsilon} \label{eq:phiOnly}\\
\nabla^2 \vec{A} + k_0^2 \vec{A} = -\mu j \label{eq:AOnly}
\end{align}

On this note, we have since discovered the work of Boyse and Paulsen, who had started to develop a nodal element based formulation by substituting the potentials into Maxwell's equations and applying the Lorenz gauge. \cite{BoyseMainPaperOnNodalPotentialFormulation,BoyseTheoretical}. Their weak formulation is missing some of the coupling terms that appear in the Lagrangian formulation with gauge fixing and the resulting strong form of the problem has no direct coupling between $\vec{A}$ and $\phi$ in the volume. However, the main issue in their original formulation is precisely the application of the boundary condition scheme suggested above, decoupling $\vec{A}$ and $\phi$. While not mentioned in the original works, a nodal implementation of the formulation will fail on non-convex domains, as noted in ref. \onlinecite{PotentialNotWorking}. Nonetheless, it is perfectly acceptable to adopt such an approach, as long as a mixed formulation is employed, with $\vec{A}$ expanded by the edge elements.

This decoupled, mixed formulation approach has been adopted by some low-frequency solvers. In these implementations the Lorenz gauge simplifies to the usual divergence free condition on $\vec{A}$ and the two components are coupled only in the sense that their excitations are related through the charge continuity equation, $\nabla \cdot \vec{J} = i \omega \rho$.\cite{LorenzGaugedMixed}  In this sense, it is not a fully four dimensional solution.

Instead, we allow the residual gauge freedom to persist by imposing boundary conditions solely through surface integrals instead of explicit Dirichlet boundary conditions. This is not a significant drawback, as surface integrals are commonly used to implement impedance or absorbing boundary conditions in any case. The natural boundary condition in our formulation corresponds to a perfect magnetic boundary. An impedance boundary condition can be imposed through the additional surface integral given by eq.~\ref{eq:surfImp}. A perfect electric boundary is imposed in the limit where the conductance, directly proportional to $\gamma$ is large. For further discussion of the boundary condition imposed by eq.~\ref{eq:surfImp} and for a definition of $\gamma$ in the context of our implementation, please see the appendix.
\begin{equation}
S_\mathrm{Z} = \gamma \int_{\partial \Omega} \left|\hat{n} \times \left(-\nabla \phi -  i\omega \vec{A} \right) \right|^2 dS \label{eq:surfImp}
\end{equation}

In contrast to the mixed formulations, the resulting solution is not a simple superposition of independent solutions for $\vec{A}$ and $\phi$, but a self-consistent solution for $\mathbf{A}$ in its entirety, as is demonstrated in section \ref{sec:Results}. This, in conjunction with the gauge fixing term which regularizes the problem, is what enables the use of nodal elements in the four-potential formulation compared to existing $\vec{A} - \phi$ formulations  \cite{LorenzGaugedMixed,H1MixedDuan,FullFITEMBaumanns, AmroucheNonSmooth,WangTimeDomainTCT}. Boyse and Paulsen arrived at a similar conclusion with their Maxwell based $\vec{A} + \phi$ formulation, implementing an impedance boundary condition in a later paper which they then demonstrated working on a 2D wedge geometry. \cite{BoyseNonConvex} 

Finally, we conclude this section by introducing a new coefficient for the gauge fixing term:
\begin{equation}
\alpha = \frac{1}{\xi} \label{adef}
\end{equation}
This is both for the sake of brevity as the coefficient of the gauge fixing terms is referred to often in the implementation and results section, and also as we are interested in plotting solution properties as a function of $\alpha$ near 0.

\subsection{Gauss' Law}
While the ability to use nodal elements is a nice benefit, the primary motivation for our adoption of the Lagrangian formulation is the fully general treatment of the source terms provided. In the appendix, we provide the full derivation of the strong form corresponding to eq.~\ref{eq:THaction} or, in the language of variational calculus, the Euler-Lagrange equations resulting from the variation of the Lagrangian. Equations \ref{eq:GL} and \ref{eq:AL} give the resulting equations imposed in the volume (there are also surface terms which are provided in the appendix). 
\begin{align}
\alpha n^2 k_0^2 \phi+ \nabla^2 \phi - i k_0 (\alpha-1) \nabla \cdot \vec{A} &= -\frac{\rho}{\epsilon}  \label{eq:GL} \\
n^2 k_0^2 \vec{A} +\nabla^2 \vec{A} + (\alpha -1) \nabla (\nabla \cdot \vec{A} + i \frac{\omega}{c^2} \phi) &= -\mu \vec{J} \label{eq:AL}
\end{align}
Equation \ref{eq:GL} is the result of the variation with respect to $\phi$ while eq. \ref{eq:AL} arises through the variation with respect to $\vec{A}$. With the Lorenz gauge implicitly imposed through the gauge fixing term, regardless of the residual gauge these equations reduce to Gauss' law and Ampere's law in this gauge. Thus, unlike in eq.~\ref{eq:CC}, both equations are independently satisfied by the solution which minimizes eq.~\ref{eq:THaction}, even in the case where discrete charge conservation is not guaranteed.

Not only does the solution explicitly satisfy Gauss' law, but the use of the nodal elements means it can do so element-wise as well as globally. This is in contrast to the lowest order edge elements, which are divergence-free within each element: any non-zero divergence in the fields arises only through discontinuities in the normal component of the field between elements of the mesh. Furthermore, there is significant flexibility offered by the fact that both $\rho$ and $\vec{J}$ can be used to drive the fields.

Both of these features are beneficial in modeling problems with significant space charge, whether for low-frequency applications where $\rho$ becomes important in the static limit or, as in our motivation for pursuing this approach, in the modeling of high frequency power sources where time harmonic components of the space charge contribute strongly to the fields even at high frequencies.

\section{Computational Implementation}
\label{sec:Implementation}
As a proof of concept, we have implemented this formulation for 2.5D azimuthally symmetric fields, solving on a 2D mesh and accounting for the azimuthal dependance of the fields, of the form $e^{i m \theta}$, a-priori. A few unique challenges arise in the implementation of the Lagrangian finite element formulation. The issue of enforcing boundary conditions through surface integrals instead of having the option of explicitly setting Dirichlet boundary conditions was discussed in section \ref{GIsection}. The other significant difference relative to the curl-curl formulation is the presence of terms linear in $k_0$.  

In finite element electromagnetic analysis, there are two types of problems which are of interest: eigenmode analysis and driven problems. In driven problems, the driving frequency is known so that only the fields need to be computed. In the eigenmode analysis, the resonant frequencies (eigenvalues) and corresponding four-potentials (eigenvectors) are calculated. The discretized Lagrangian is composed of three finite element matrices, $\mathrm{\mathbf{M}},\mathrm{\mathbf{C}},\mathrm{\mathbf{K}}$ and the resulting matrix equation is a generalized quadratic eigenvalue problem (QEP) where we solve for $\tilde{k_0}$ , the approximate resonant frequency, and $\mathbf{a}$, the coefficients of the approximate solution over the discretized space.
\begin{equation}
\left(\mathrm{\mathbf{M}}\tilde{k_0}^2 + \mathrm{\mathbf{C}} \tilde{k_0} + \mathrm{\mathbf{K}} \right).\mathbf{a} = 0
\end{equation}

It is the coupling between $\vec{A}$ and $\phi$, appearing in the matrix $\mathrm{\mathbf{C}}$, that results in a quadratic eigenvalue problem instead of the regular generalized eigenvalue problem of the curl-curl equation. QEPs are common in finite element problems, for example in modeling damped structural resonances \cite{} and a significant body of work exists on the topic, including a comprehensive review paper \cite{TisseurQEP}. 

We implemented the sparse non-linear eigenvalue solver, NLFEAST \cite{NLFEAST}, a contour integral based solver where we constructed the kernel specifically for our QEP. Our implementation has proven robust, agreeing with the direct solver for small problem sizes where a comparison was possible, and scalable up to matrix sizes on the order of 1E6 (we did not test beyond this as for a 2D mesh, this is a very dense mesh). The condition number of the eigenvalues, as defined in ref. \onlinecite{TisseurQEP}, are reasonable and uncorrelated to problem size. The conditioning does depend weakly on the gauge fixing term and the need for the gauge fixing term becomes immediately clear from the singularity in the condition number when it is not included, as will be shown in sec.~\ref{sec:Results}.

The driven problem employs the same $\mathrm{\mathbf{M}},\mathrm{\mathbf{C}}$ and $\mathrm{\mathbf{K}}$ matrices, but $\tilde{k_0}$ is set by the frequency of the driving source terms and $\mathbf{a}$ is determined by solving the resulting linear system.
\begin{equation}
\left(\mathrm{\mathbf{M}} \tilde{k_0}^2 + \mathrm{\mathbf{C}} \tilde{k_0} + \mathrm{\mathbf{K}} \right).\mathbf{a} = \mathrm{\mathbf{j}}
\end{equation}

For this, we use the Intel Math Kernel Libraries, and in particular, the PARDISO solver. We have tried direct and iterative solvers and found both to be equally effective for the moderate problem sizes we have been working with so far. 

Future work will look to scale the implementation to 3D meshes and thus much large matrices. Here, a more advanced solver and the employment of a preconditionner will likely be beneficial. We expect NLFEAST or a similar contour integral solver will still be the optimal choice for the eigenmode analysis. 

\section{Numerical Results and Benchmarking}
\label{sec:Results}
We have benchmarked the Lagrangian formulation with respect to the edge element curl-curl formulation over a broad range of examples. The following subsections focus particularly on numerical results demonstrating the accuracy, robustness and flexibility of this formulation. 

The examples shown are azimuthally symmetric, solved in a cylindrical coordinate system $(r,\theta,z)$ with an azimuthal dependence of the form $e^{i m \theta}$, as given by eq.~\ref{eq:mdep}. As this dependence is known a-priori these modes can be solved on a 2D mesh with $\theta$ out of plane. In the following, all figures of mode profiles are thus cross sectional views of the full structure in the ($z,r$) plane.
\begin{equation}
\vec{\mathcal{E}}(\vec{r},t) = \mathrm{Re}[\vec{E}(\vec{r}) e^{i \omega t+i m \theta}] \label{eq:mdep}
\end{equation}

For monopole modes ($m=0$) the fields split into modes which can be represented by $A_\theta$ alone (transverse electric or TE), or as a combination of $A_z, A_r$, and $\phi$ (transverse magnetic or TM). The TE modes are not susceptible to the challenges discussed previously and are already often solved using nodal basis functions so we focus only on TM modes for $m=0$. To fully prove the suitability of the Lagrangian formulation, particularly in regards to eventual extension to a full 3D finite element implementation, we also demonstrate some examples of dipole ($m=1$) and quadrupole modes ($m=2$). In this case, the problem is fully four-dimensional and all components of the four-potential couple to each other.

For comparison we used COMSOL, a commercially available multi-physics finite element software which includes an edge-element electromagnetic field solver. It is capable of solving axisymmetric in-plane fields on a 2D mesh, allowing for a comparison with our computational implementation in terms of accuracy and problem size. We refer the reader to the COMSOL user manual for exact implementation details\cite{COMSOL}. While our implementation uses nodal Lagrange elements and COMSOL is using edge elements, in both cases the elements are second order.

Finally, in the convergence plots that follow, we define the error as follows: for the frequency, the error is computed as $\Delta f = \frac{f-f_\mathrm{theor}}{f_\mathrm{theor}}$ if a theoretical solution exists, or for the ridge waveguide, by the frequency of the problem on a finer mesh than those plotted. For the fields, we calculate the $S_0$ (Sobolev Zero) norm of the field error over the entire problem domain: $\Delta E = \left| \frac{E-E_\mathrm{theor}}{E_\mathrm{theor}} \right|_{S0}$. To keep the plots legible, instead of showing the error of all six fields, we use the averaged error norm, $\Delta F = \frac{1}{6}(|\Delta E_r|_{S0} + |\Delta E_z|_{S0}+...)$.

\subsection{Accuracy}
\begin{figure}
\begin{subfigure}{0.5\columnwidth}
\includegraphics[width=\columnwidth]{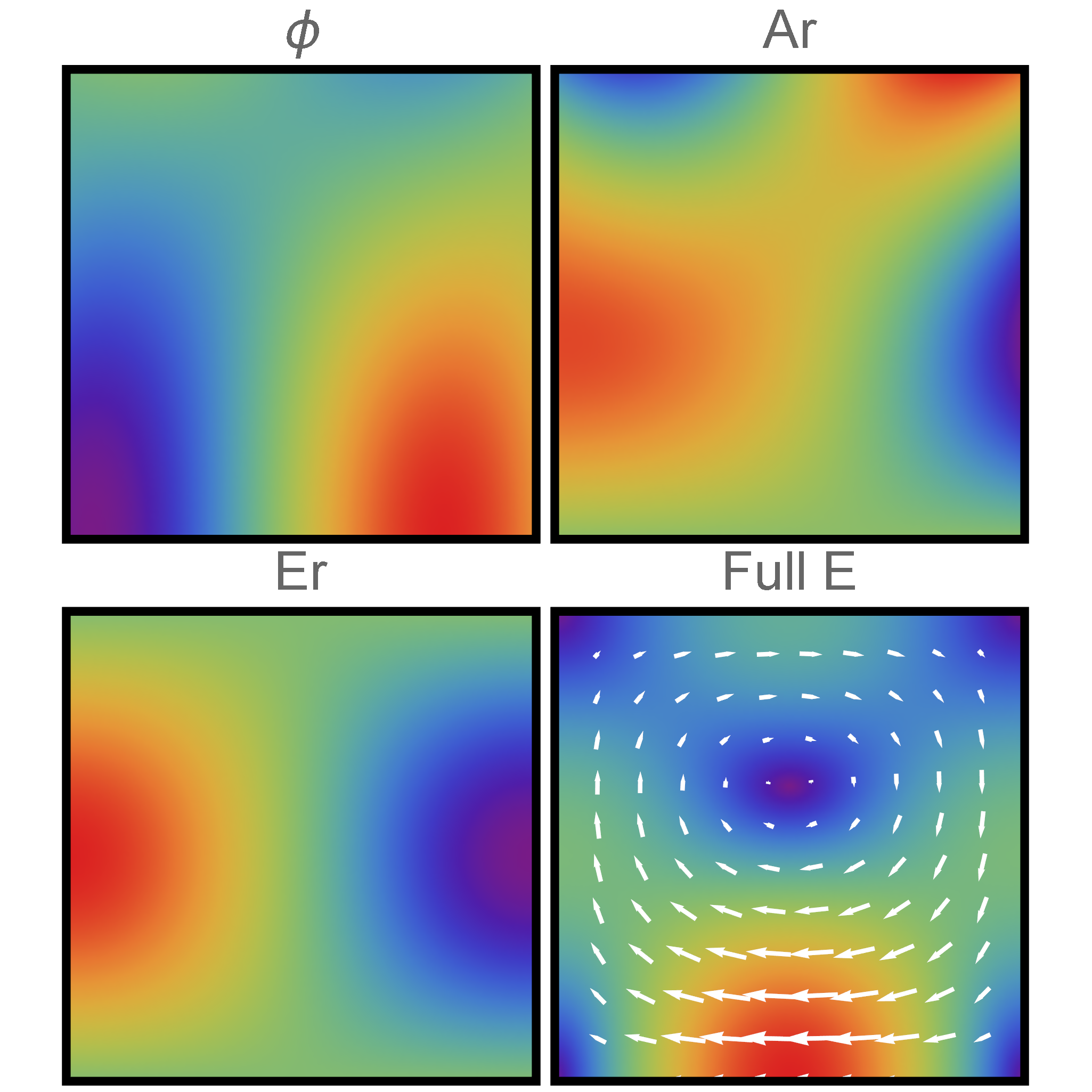}
\caption{$\alpha= 1$}
\label{fig:Pillbox1}
\end{subfigure}\begin{subfigure}{0.5\columnwidth}
\includegraphics[width=\columnwidth]{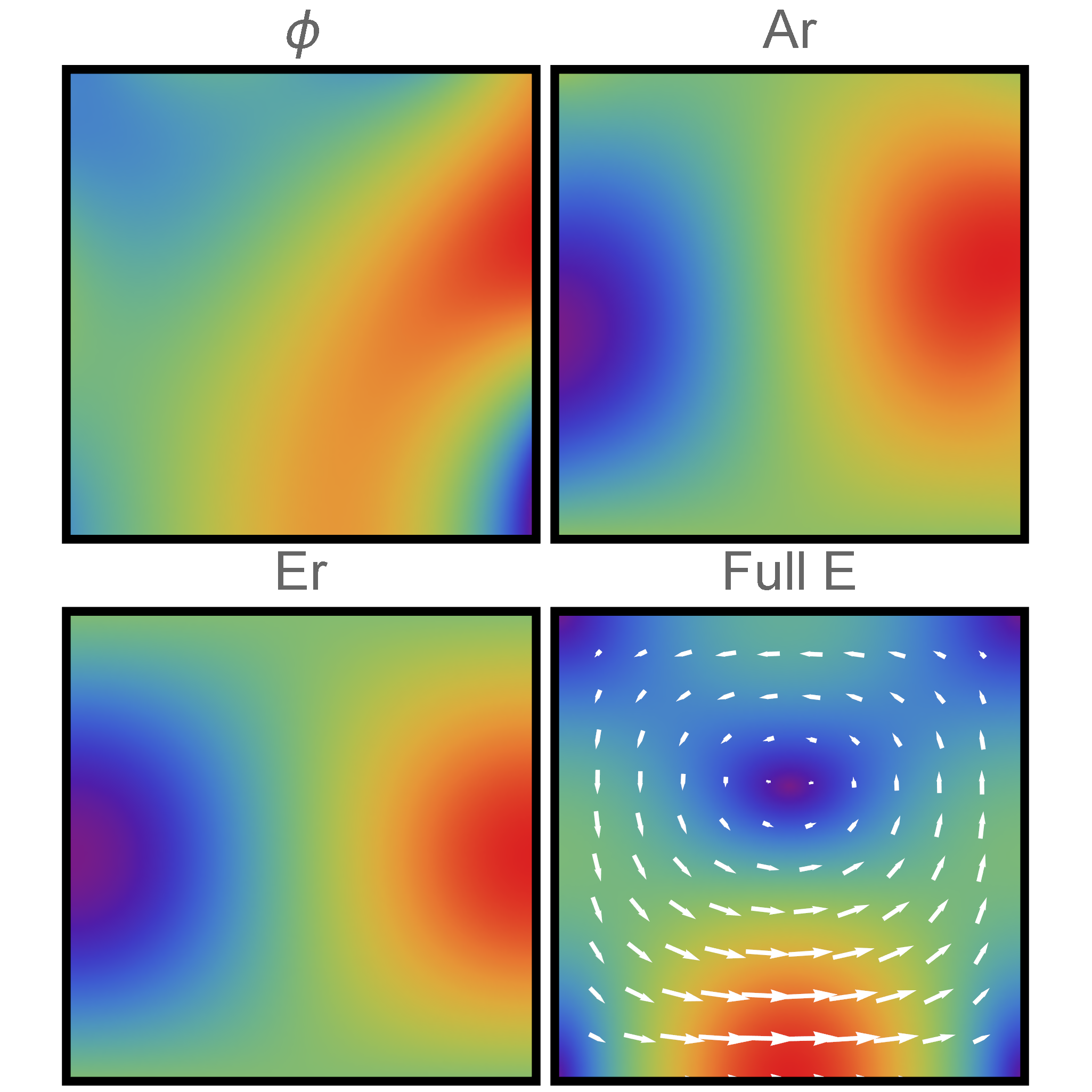}
\caption{$\alpha = -1$}
\label{fig:Pillbox2}
\end{subfigure}
\caption{Finite element solution for the TM$_{011}$ mode of a cylindrical cavity with perfect magnetic boundary for two different values of $\alpha$. The mesh used to compute the solution is overlayed in the top left figure.} \label{fig:PillboxComp}
\end{figure}

The cylindrical pillbox cavity is a good initial test case as results can be compared to the analytical solution. Figure \ref{fig:PillboxComp} shows the cross sectional problem geometry and the mode profiles for the TM$_{011}$ mode with a perfect magnetic boundary condition on the walls. Only the components of $E_r$ and the full vector field plot are shown for the sake of brevity. Two distinct solutions for the four-potential are shown, however, corresponding to different values of the gauge fixing coefficient $\alpha$ defined in eq.~\ref{adef}. Changing $\alpha$ numerically perturbs the system, producing a solution with a different residual gauge, $\psi$. Nonetheless, the resonant frequencies and fields calculated from the different solutions for the four-potential correspond to the same mode. 
\begin{figure}
\includegraphics[width=\columnwidth]{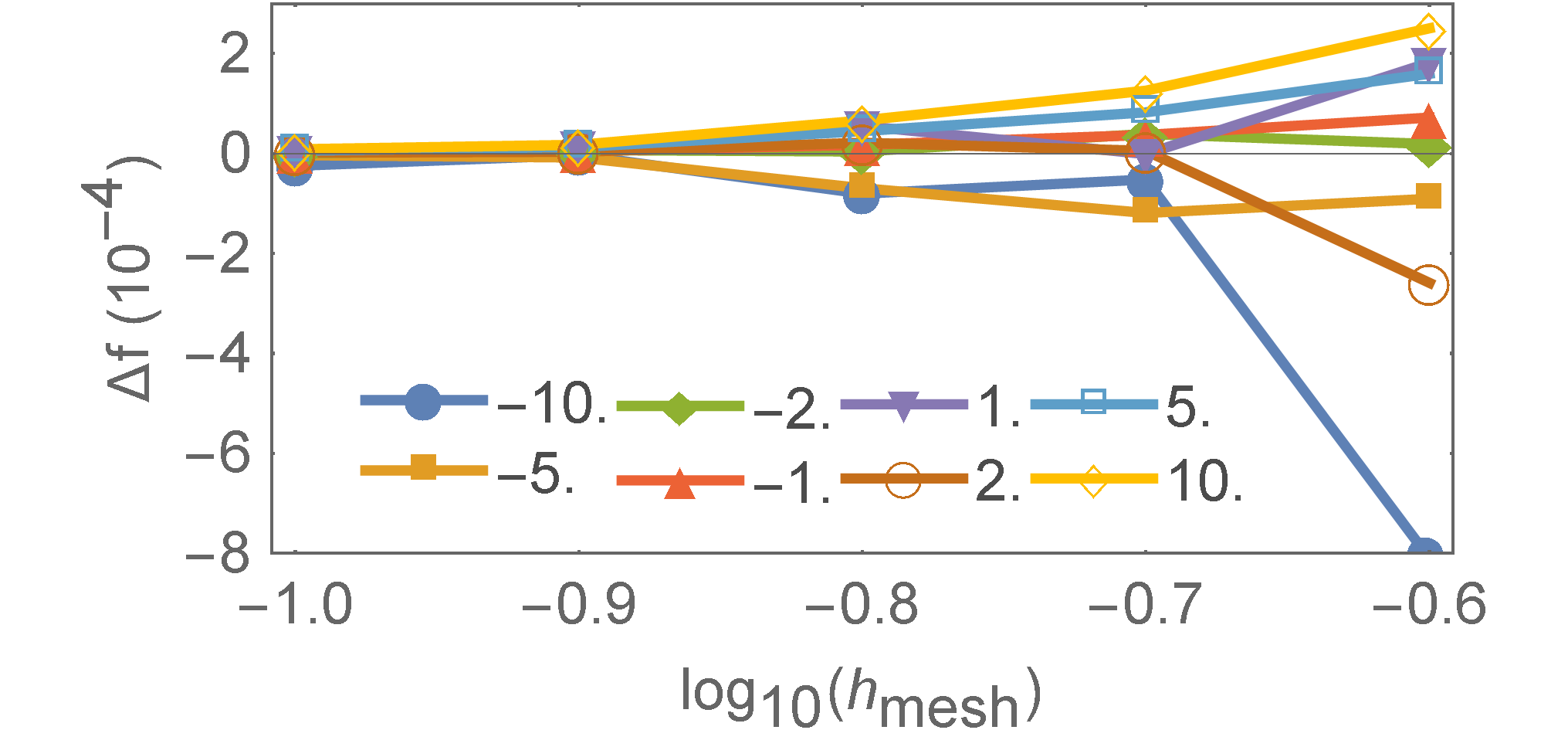}
\caption{Error in the frequency, $\Delta f = \frac{f_{FE}-f_{theor}}{f_{theor}}$ for various values of $\alpha$. The variation in solved frequency decreases as the mesh is refined.}
\label{fig:freqPillbox}
\end{figure} 
\begin{figure}
\includegraphics[width=\columnwidth]{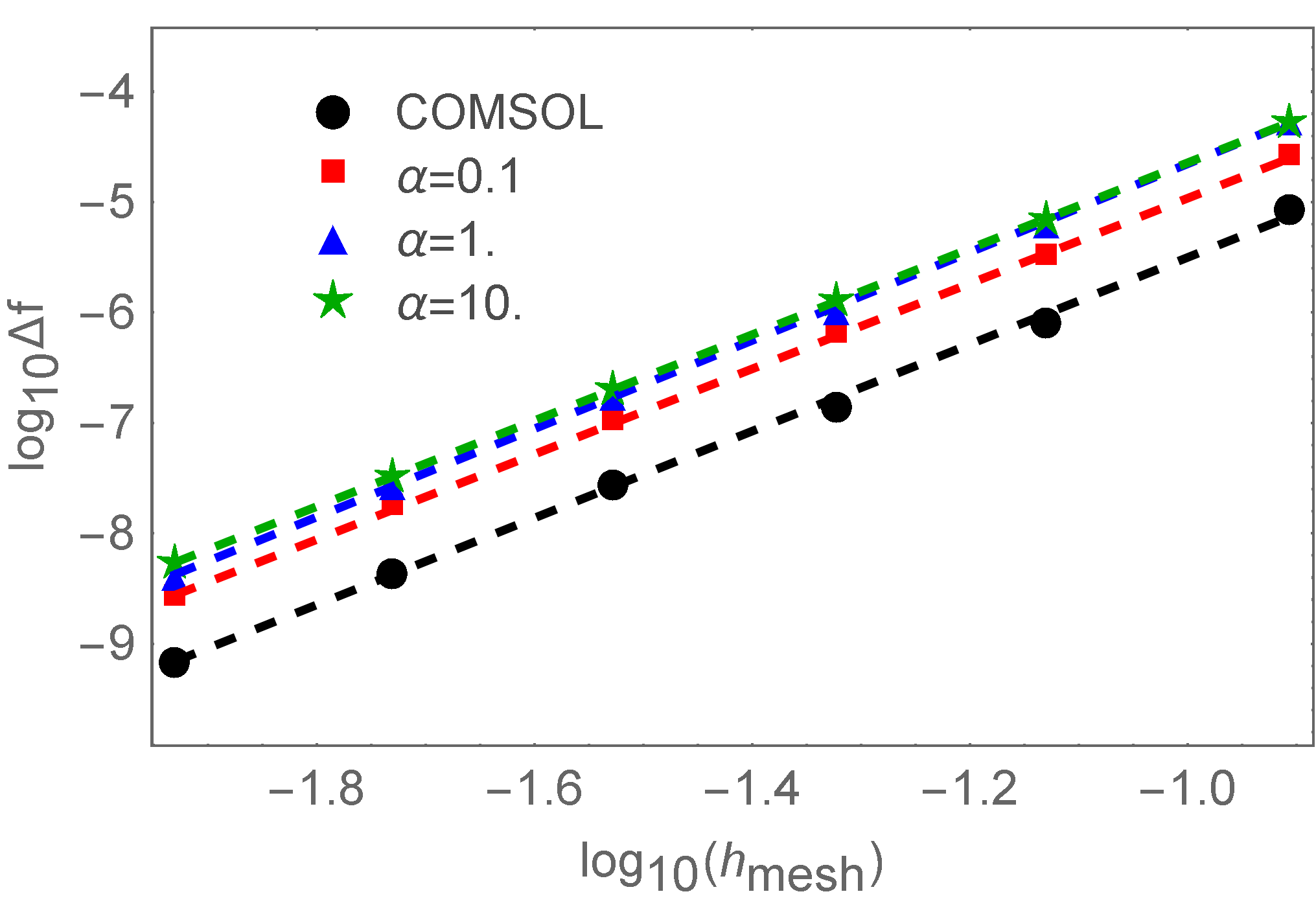}
\caption{Convergence of the frequency for the mode in fig.~\ref{fig:PillboxComp} with mesh size, $h_\mathrm{mesh}$. The slopes of the linear fits are 3.93 (COMSOL), 3.86 ($\alpha=0.1$), 4.00 ($\alpha =1$ and 3.89 ($\alpha=10$).}
\label{fig:convergencePillbox}
\end{figure}

An interesting consequence of calculating different $\mathbf{A}$ for the same mode is that the numerical error is different in each case, as demonstrated by fig.~\ref{fig:freqPillbox}. As the mesh is refined, all solutions converge to the same frequency and fields. Plotting this convergence, now for only a few values of $\alpha$, fig. \ref{fig:convergencePillbox} demonstrates similar convergence characteristics for both the nodal and edge elements. The slopes of the linear fits match that predicted from theory for second order elements, converging as O($h^4$) where $h$ is the maximum mesh edge length.

\begin{figure}
\includegraphics[width=\columnwidth]{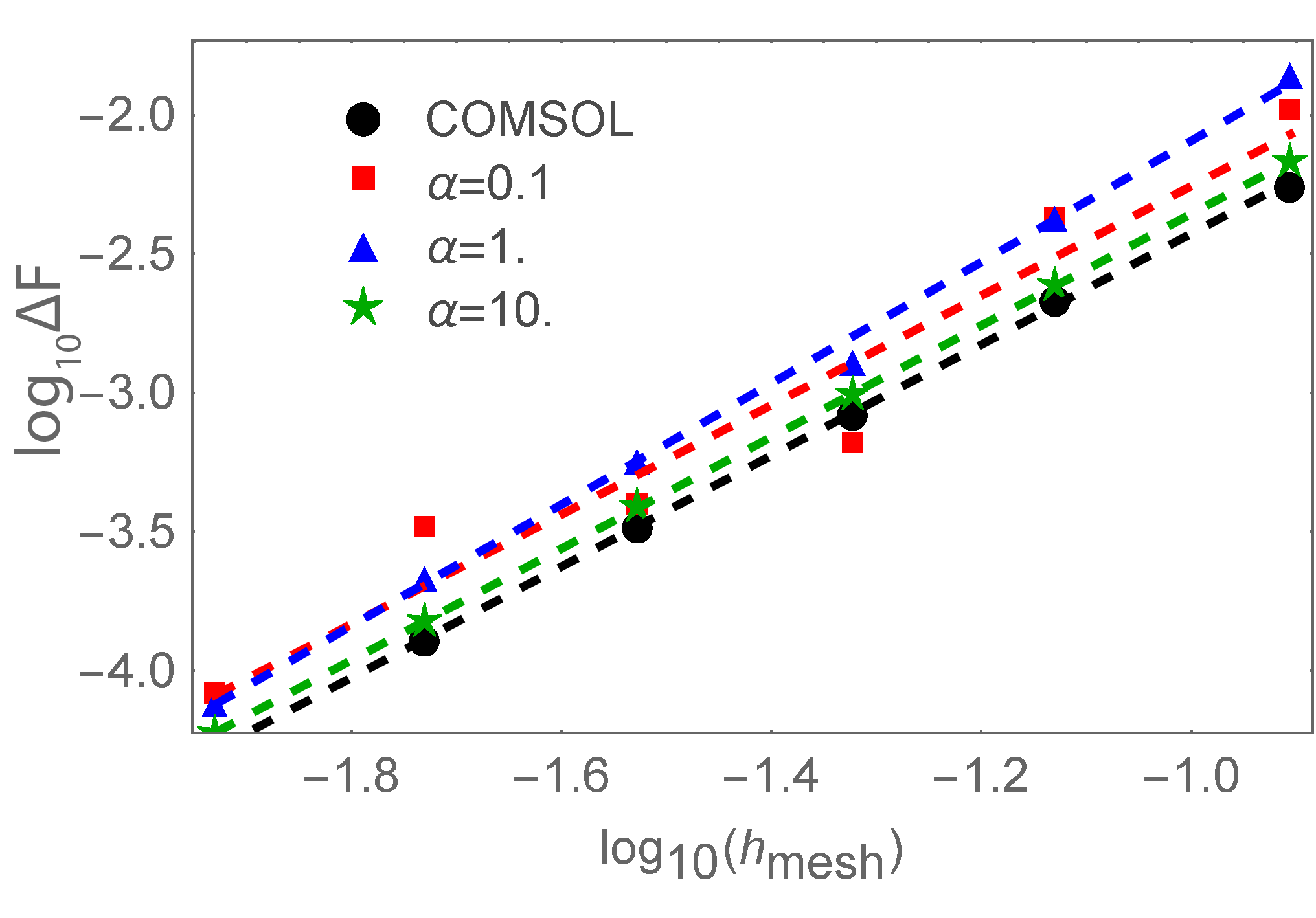}
\caption{Convergence of the fields, using the average error of $E_r$, $E_z$ and $H_\theta$, for the TM$_011$ mode with mesh size, $h_\mathrm{mesh}$. The slopes of the linear fits are 2.00 (COMSOL), 1.97 ($\alpha=0.1$), 2.18 ($\alpha=1$) and 2.01 ($\alpha=10$).}
\label{fig:convergenceFieldsPillbox}
\end{figure}

A possible downside in solving for the four-potential is that the desired end results are the electromagnetic fields, not the potentials. As the fields are obtained through derivatives of the potential, they are not expected to converge at the same rate as the solution itself. This is also an issue with the curl-curl formulation, as the magnetic field must be calculated from the solution for the electric field or vice-versa. There are methods to resolve or mitigate this issue, for example the superconvergent patch recovery technique often employed to compute stress in structural mechanics problems.\cite{spr} However, here we take the simplest approach, taking derivatives of the second order basis functions to compute the fields at the mesh nodes, which still produces comparable results with those computed by COMSOL. The convergence of the fields for the mode in fig.~\ref{fig:PillboxComp} are given by fig.~\ref{fig:convergenceFieldsPillbox}.

\subsection{Flexibility}
\begin{figure}
\includegraphics[width=0.8\columnwidth]{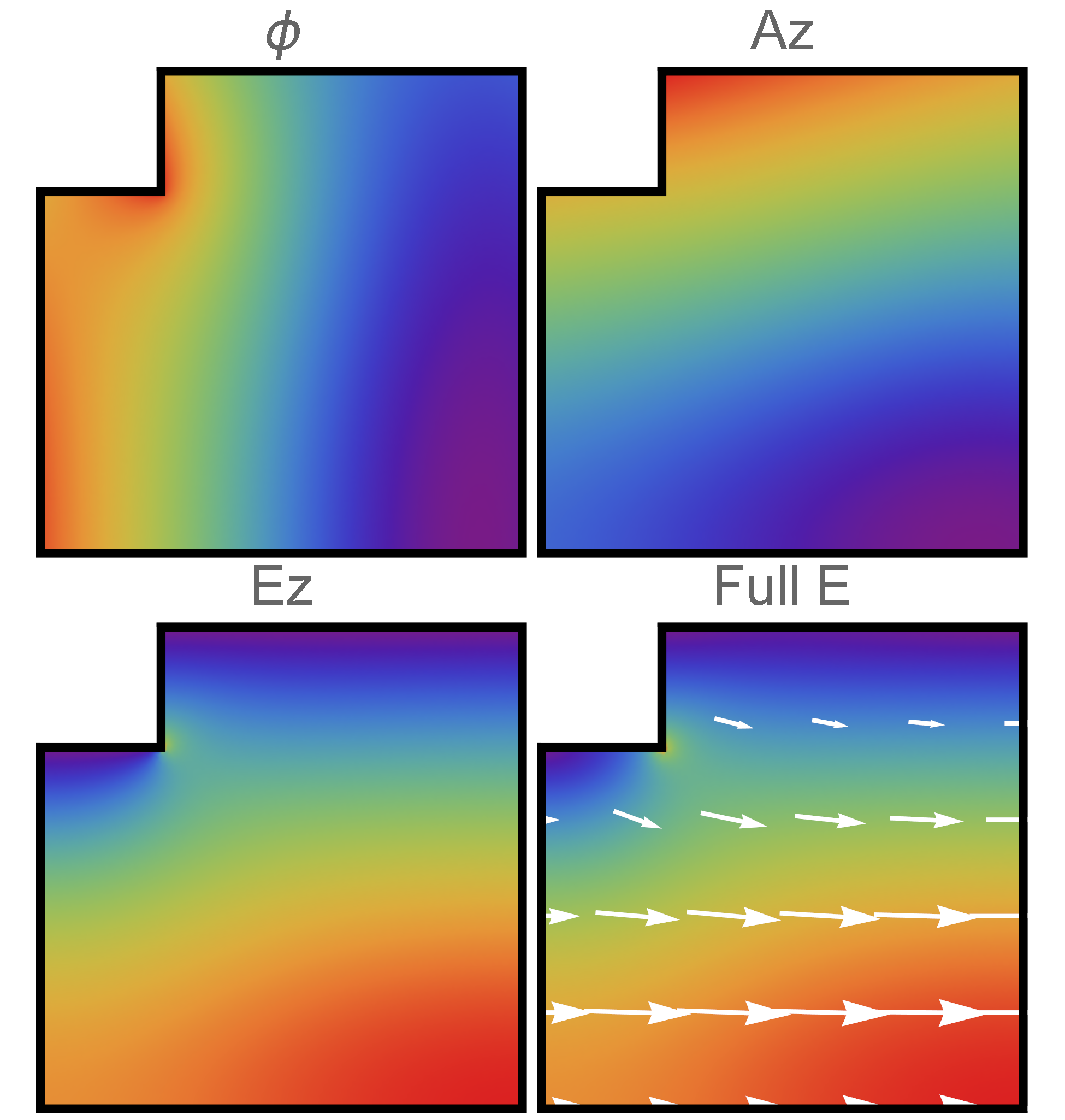}
\caption{FE solution for the fundamental TM mode of a notched pillbox cavity, f=120.0 MHz. $\phi$, $A_r$ are continuous but the singularity is captured in the computed fields through $\nabla \phi$.}
\label{fig:RidgedWaveguide}
\end{figure}

\begin{figure}
\includegraphics[width=0.8\columnwidth]{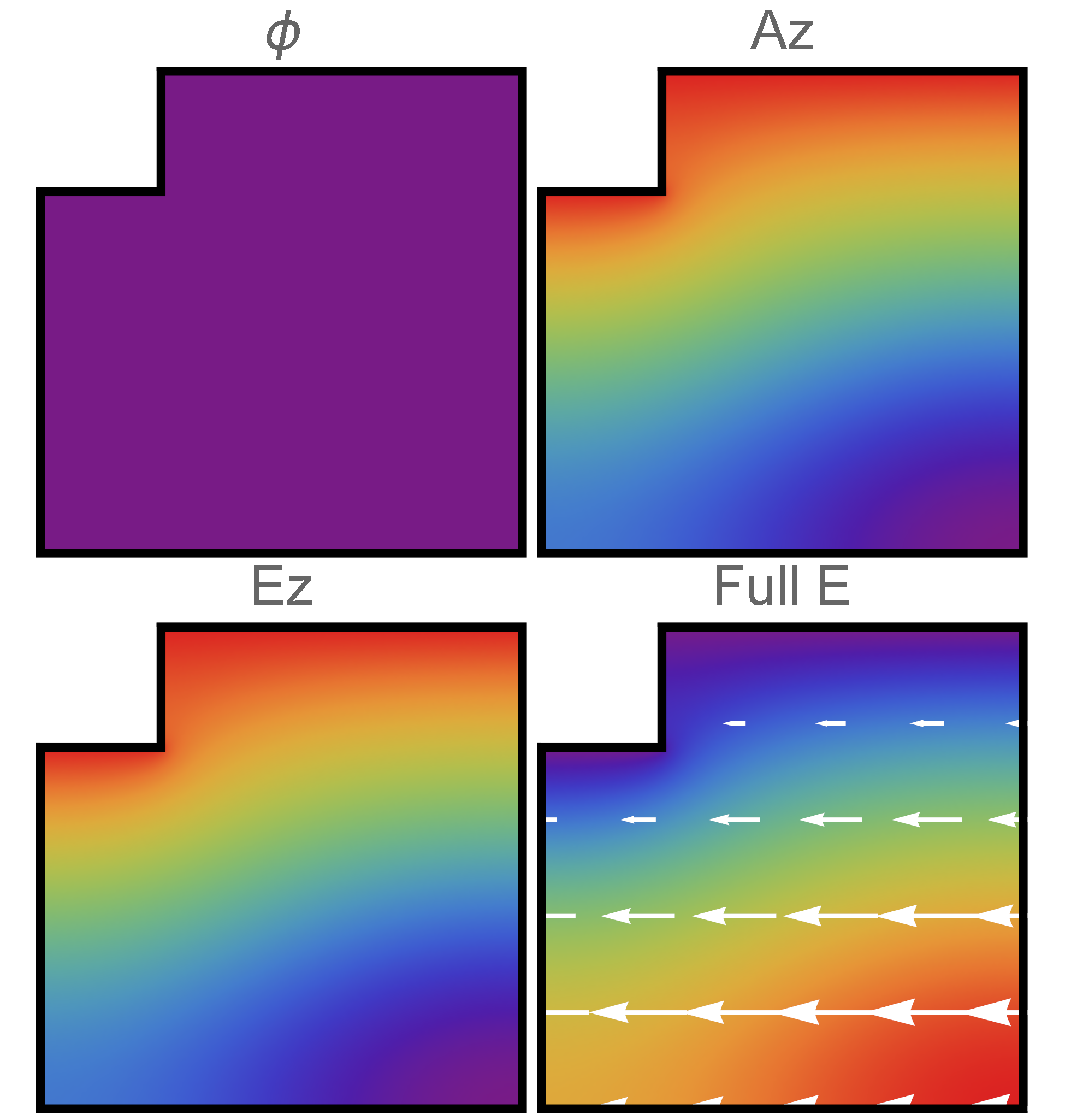}
\caption{FE solution for the fundamental TM mode of a notched pillbox cavity where $\phi$ is set to zero, f=127.1 MHz. $\vec{A}$ alone does not resolve the singularity.}
\label{fig:RidgedWaveguideNoPhi}
\end{figure}

Moving on to problems where nodal element based solvers using the conventional curl-curl equation fail, fig.~\ref{fig:RidgedWaveguide} plots the solution for a notched pillbox cavity with a perfect electric boundary. There is a singularity in the fields on the corner which conventional nodal field solvers cannot resolve, converging to the incorrect solution even as the mesh is refined. While the solution for the four-potential is continuous, the discontinuity in the fields at the notch is fully captured by $\nabla \phi$ in the Lagrangian formulation, as demonstrated in the figure. In this case, the frequency computed by COMSOL and the four-potential formulation is 120.0 MHz. If instead, we set $\phi=0$ either on the boundary or the entire volume, decoupling the four-potential, we find that instead the frequency computed is 127.1 MHz. The field profile for this (incorrect) mode is shown in fig.~\ref{fig:RidgedWaveguideNoPhi}, now with no singularity at the re-entrant corner. 

In fig.~\ref{fig:convergenceRidgeWaveguide}, the convergence of the frequency is plotted as a function of the number of degrees of freedom solved for. Instead of plotting as a function of mesh size, where we do not expect to obtain a theoretical rate of convergence due to the singularity in any case, we plot as a function of problem size to illustrate another perhaps counter-intuitive result.  The absolute accuracy relative to problem size is comparable despite the additional degree of freedom used in the four-potential formulation. This is because edge elements require roughly twice as many degrees of freedom as nodal elements for the same convergence order\cite{GMurAdvDis,FEMtextbook}. This is in part due to the additional degrees of freedom per mesh element and in part because there are many more edges than nodes in a mesh.
\begin{figure}
\includegraphics[width=\columnwidth]{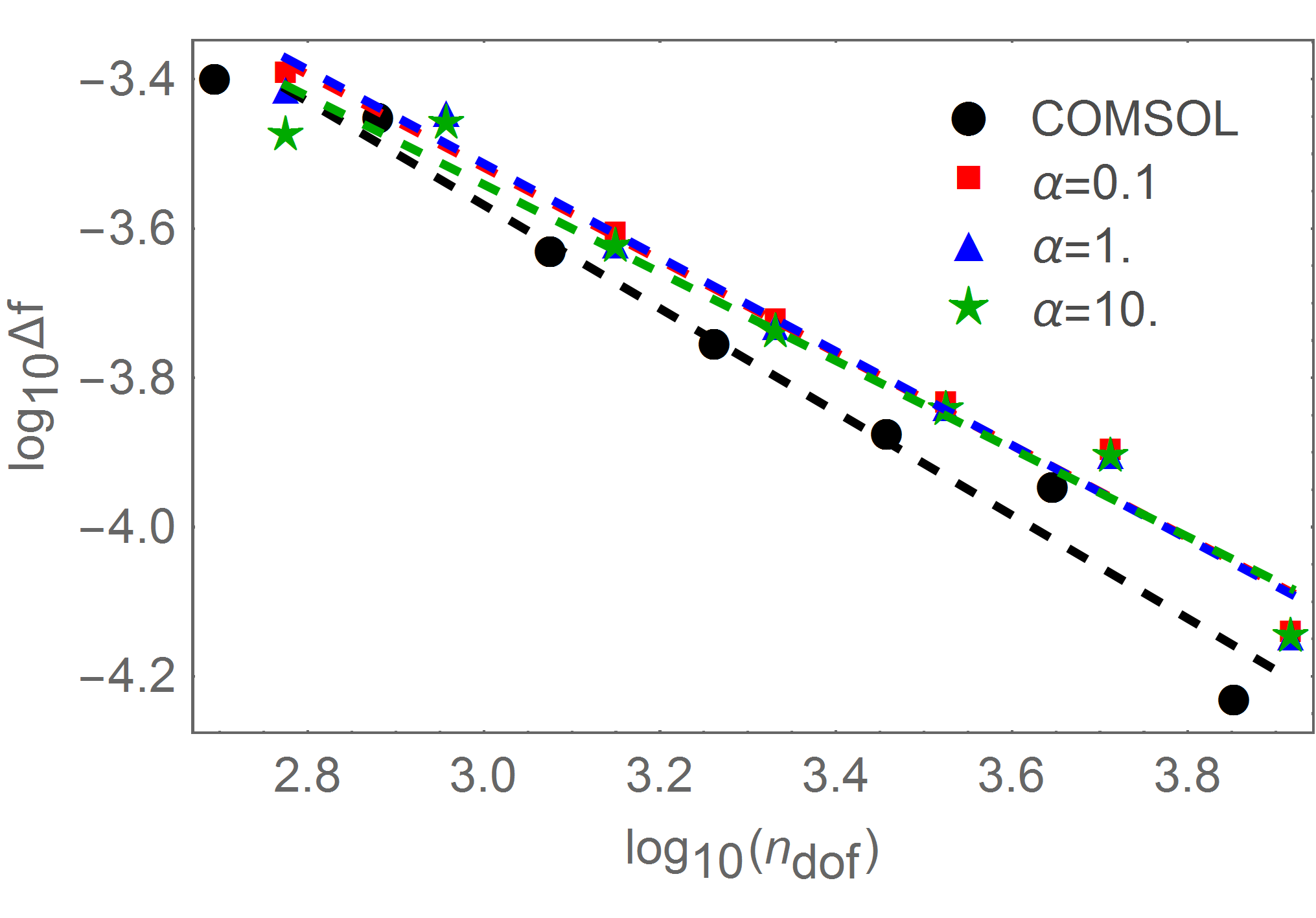}
\caption{Convergence of the frequency, $f_0$, for the mode in fig.~\ref{fig:RidgedWaveguide} with problem size, $n_{DOF}$.}
\label{fig:convergenceRidgeWaveguide}
\end{figure}
\begin{figure}
\includegraphics[width=\columnwidth]{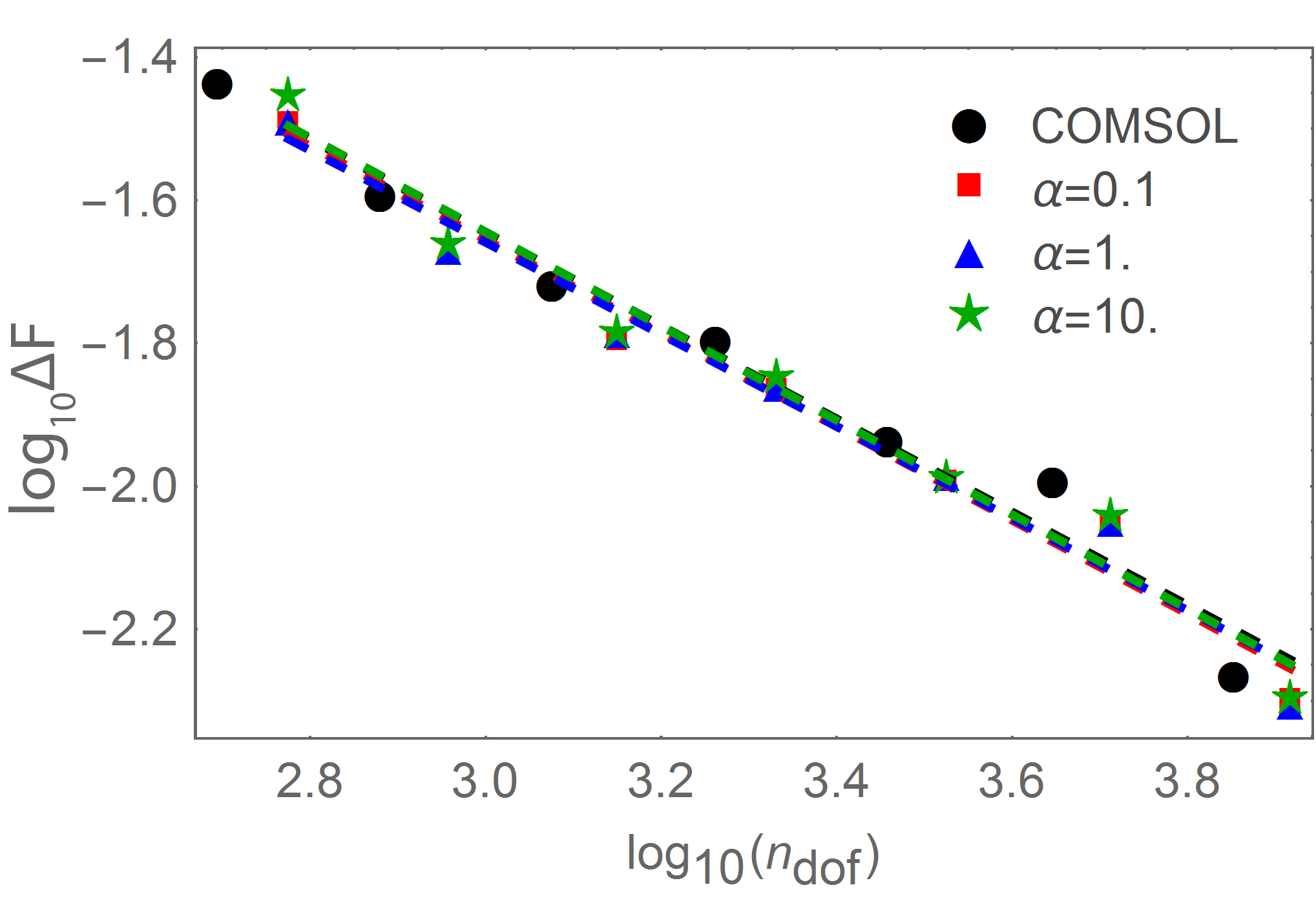}
\caption{Convergence of the fields (average error over all field components), $f_0$, for the mode in fig.~\ref{fig:RidgedWaveguide} with problem size, $n_{DOF}$.}
\label{fig:convergenceRidgeWaveguide}
\end{figure}
Spherical cavities can also be modeled in 2.5D, presenting another example with re-entrant corners (when approximated as a polygon) but one with a theoretical solution to which we can compare. As we have not yet implemented curvilinear or isoparametric elements, the convergence rate in this case is dominated by the extent to which the curved boundary is approximated by a polygon. The results shown are for the TM$_{331}$ mode but note that in this case, TM refers to transverse magnetic with respect to $\rho= \sqrt{r^2+z^2}$, the convention for spherical cavities, and not with respect to $\theta$ so all components of the four-potential must be solved for.
\begin{figure}
\includegraphics[width=0.8\columnwidth]{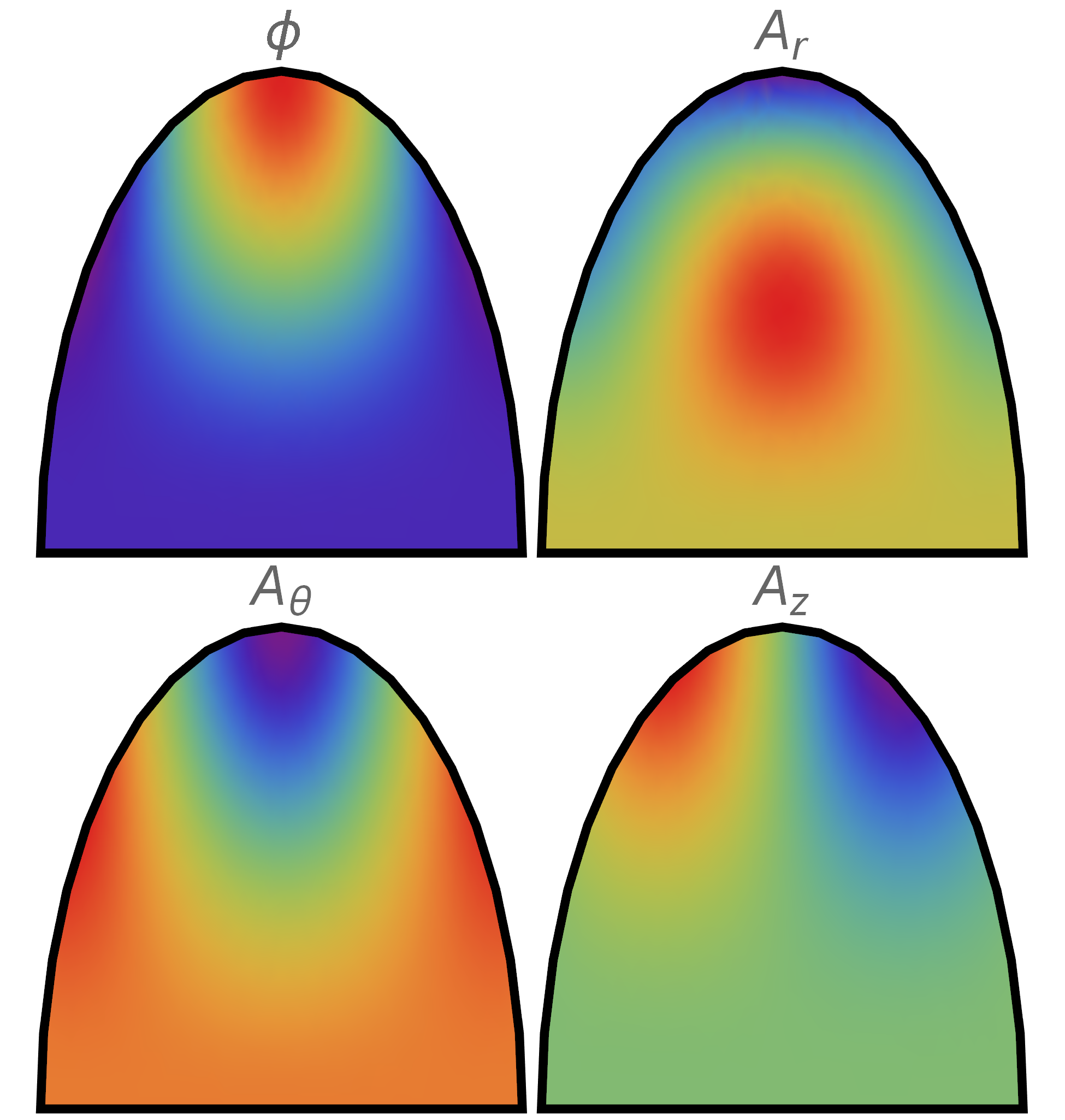}
\caption{Mode profile for the four-potential for the TM$_{331}$ mode of a spherical cavity with an impedance boundary.}
\label{fig:sphereZDipole}
\end{figure} 
\begin{figure}
\includegraphics[width=\columnwidth]{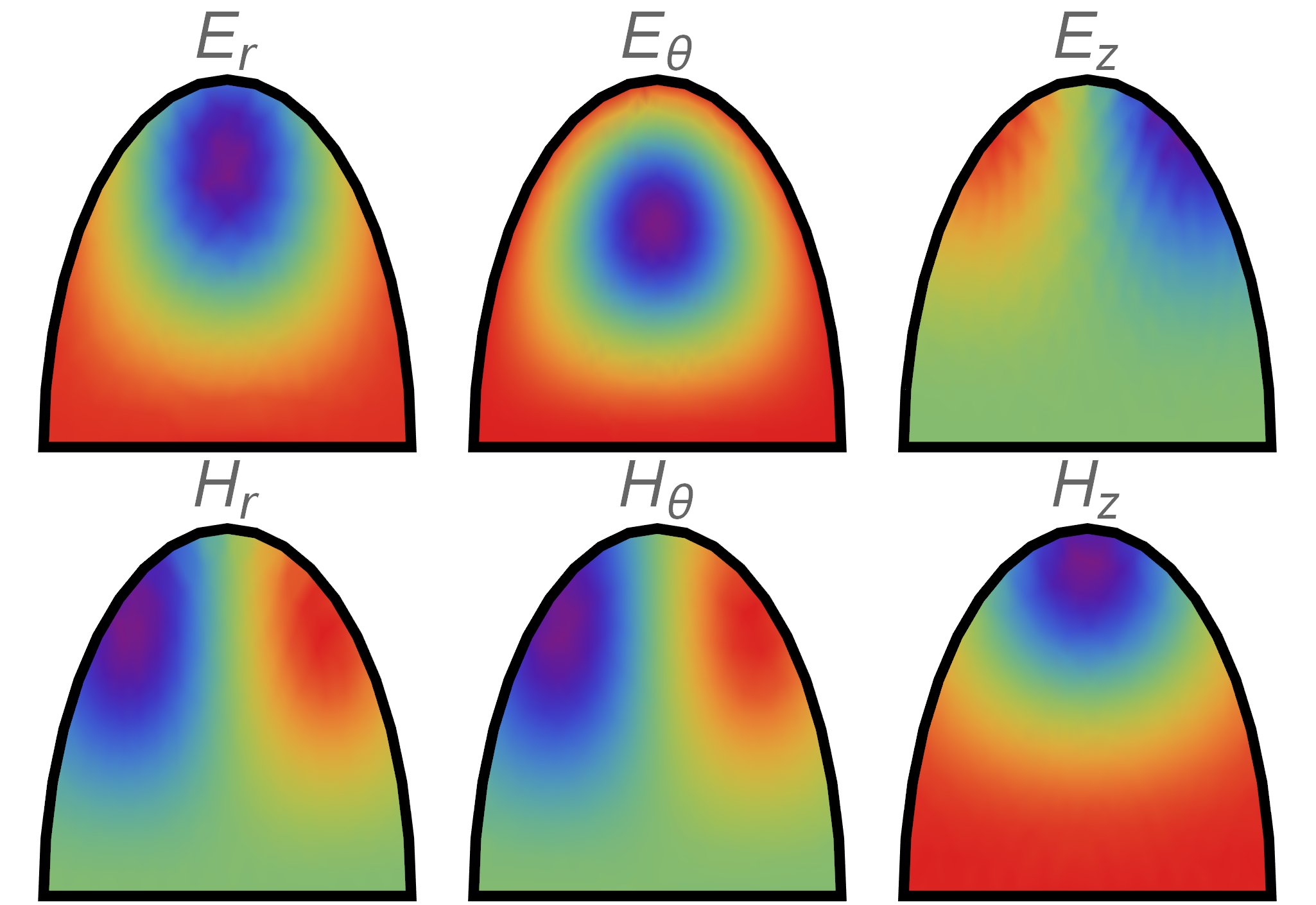}
\caption{Computed fields for the mode in fig~\ref{fig:sphereZDipole}}
\label{fig:SphereZDipoleF}
\end{figure} 
\begin{figure}
\includegraphics[width=\columnwidth]{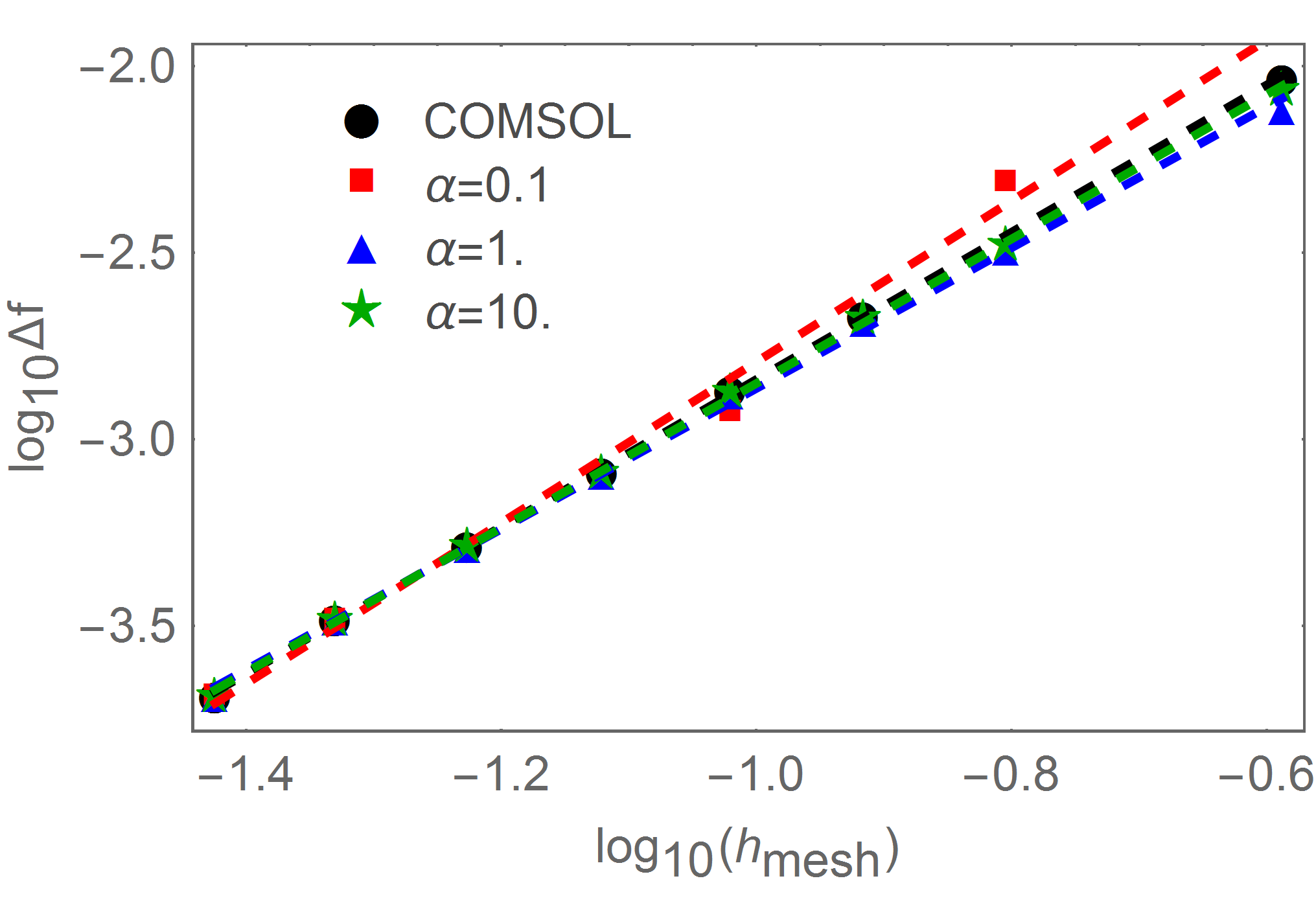}
\caption{Convergence of the frequency for the mode in fig.~\ref{fig:sphereZDipole} with mesh size, $h_{mesh}$. The slopes of the linear fits are 1.97 (COMSOL), 2.16 ($\alpha=0.1$), 1.89 ($\alpha=1$) and 1.93 ($\alpha=10$).}
\label{fig:convergenceSphereDipole}
\end{figure} 
\begin{figure}
\includegraphics[width=\columnwidth]{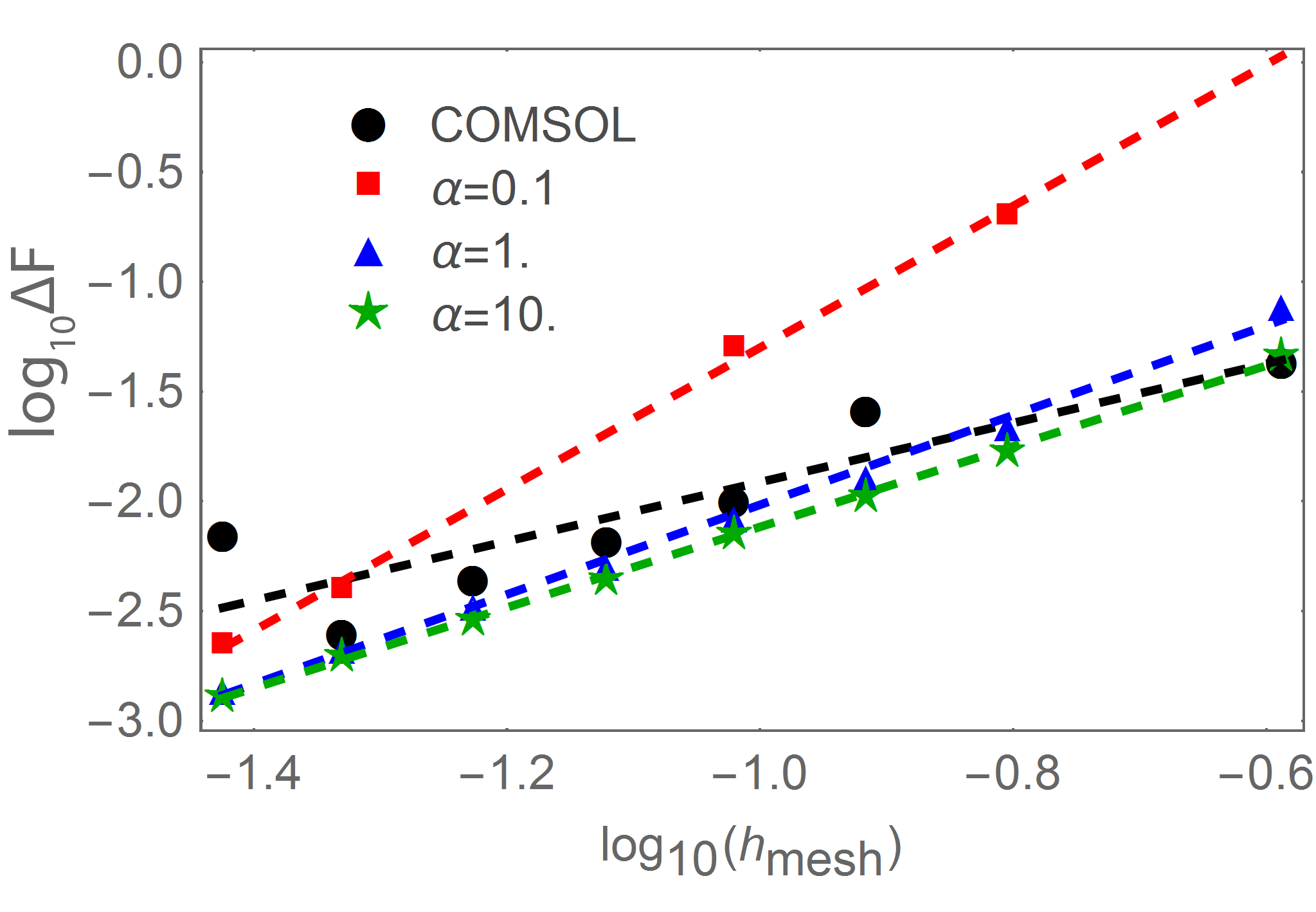}
\caption{Convergence of the fields (average error over all field components) for the mode in fig.~\ref{fig:sphereZDipole} with mesh size, $h_{mesh}$. The slopes of the linear fits are 1.35 (COMSOL), 3.23 ($\alpha=0.1$), 2.04 ($\alpha=1$) and 1.84 ($\alpha=10$).}
\label{fig:convergenceESphereDipole}
\end{figure}

\begin{figure}
\includegraphics[width= \columnwidth]{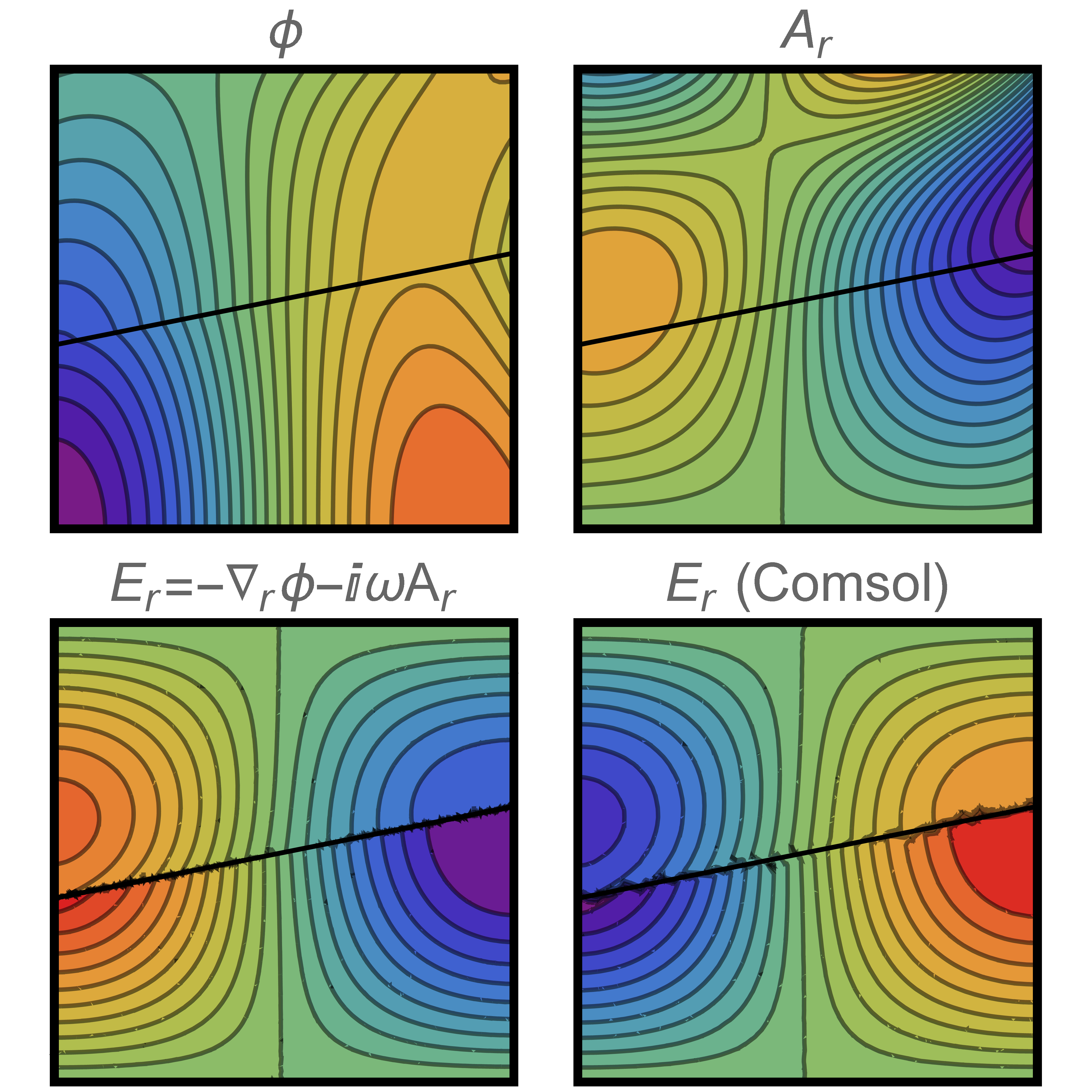}
\caption{FE solution for the fundamental TM mode of a dielectric lined cavity with $\epsilon_r=1.5$ above the thick back line. The resonant frequency is $f_0=214.051$ MHz compared to $214.054$ MHz in COMSOL.}
\label{fig:dielFields}
\end{figure}

Finally, in addition to field singularities due to singular boundaries and re-entrant corners, the four-potential formulation can model discontinuities at material interfaces without the special treatment typically required to accommodate the jump in the normal field\cite{FEMtextbook}. Figure \ref{fig:dielFields} shows the field profile for a tapered dielectric lined cavity, for example. $\phi$ and $\vec{A}$ are continuous but $\nabla \phi$ captures the discontinuity in the fields due to the change in $\epsilon_r$. 

\subsection{Robustness}
\begin{figure}
\begin{subfigure}{\columnwidth}
\includegraphics[width=0.75\columnwidth]{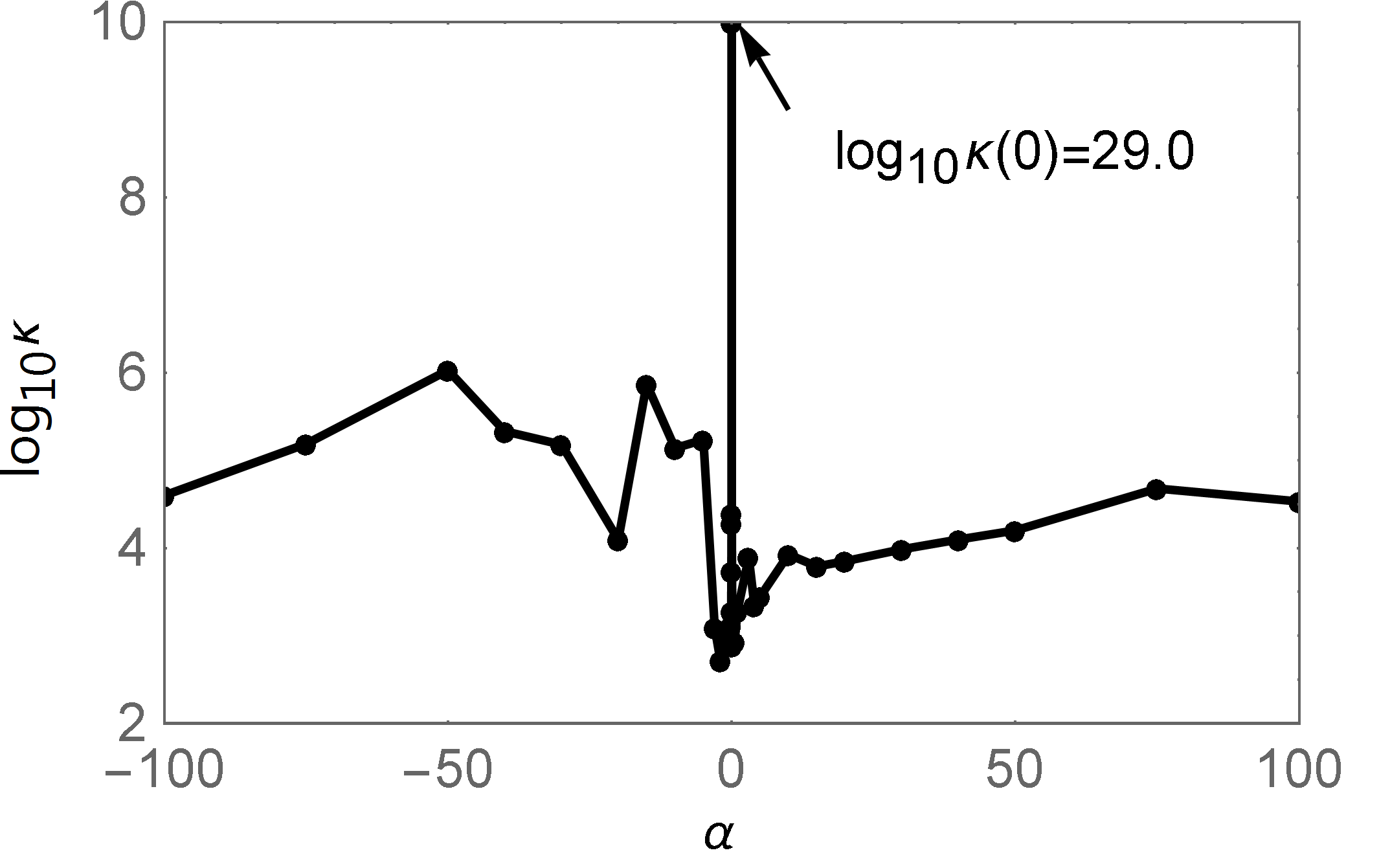}
\caption{Condition number as a function of gauge fixing.}
\label{fig:condNumberPMa}
\end{subfigure}
\begin{subfigure}{\columnwidth}
\includegraphics[width=0.75\columnwidth]{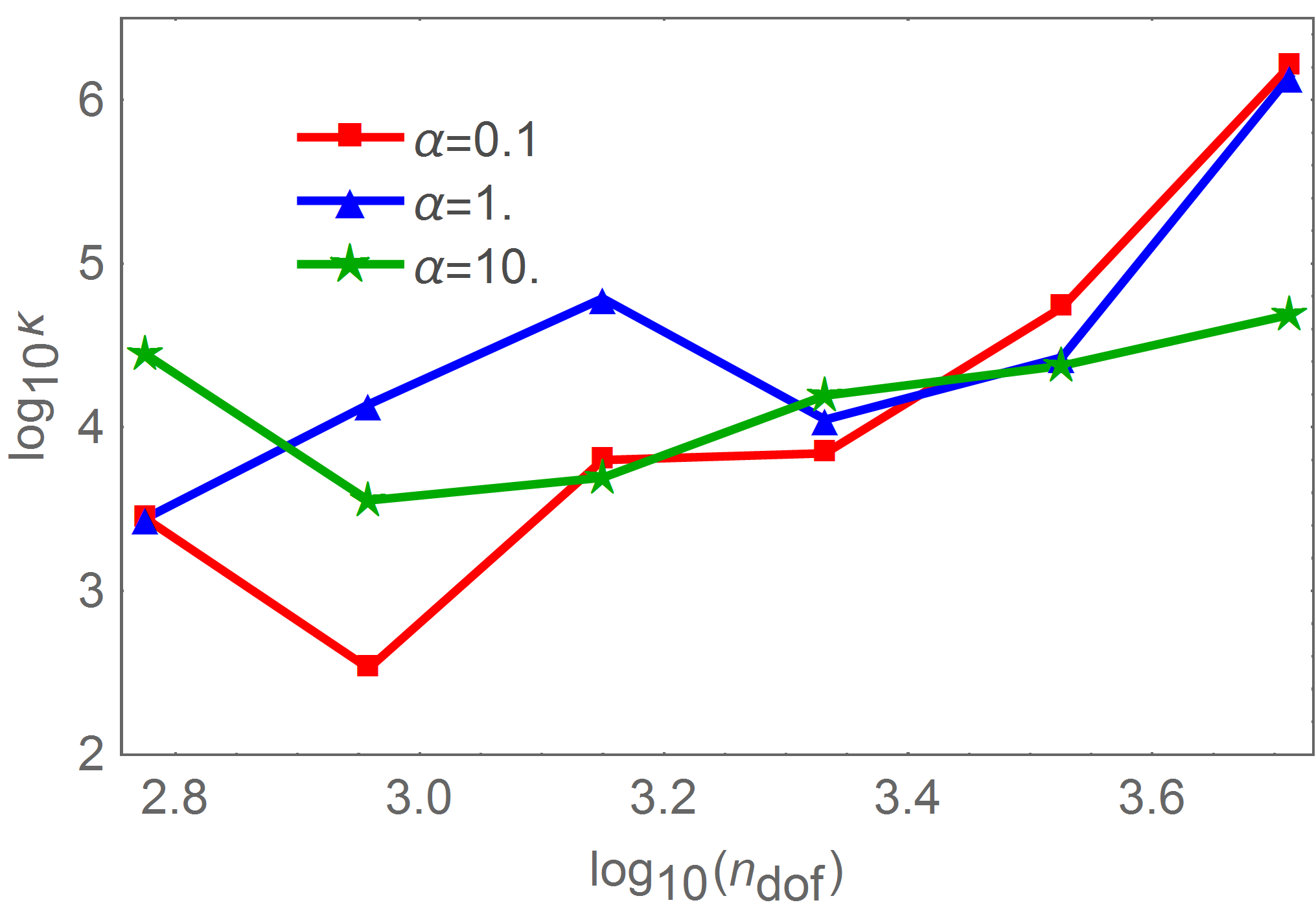}
\caption{Condition number as a function of mesh resolution.}
\label{fig:condNumberPMm}
\end{subfigure}
\caption{Condition number, $\kappa$ for the eigenvalue corresponding to the fundamental TM mode of the cavity shown in fig.~\ref{fig:RidgedWaveguide} with perfect magnetic boundary.}
\label{fig:condNumbPM} 
\end{figure}

\begin{figure}
\begin{subfigure}{\columnwidth}
\includegraphics[width=0.75\columnwidth]{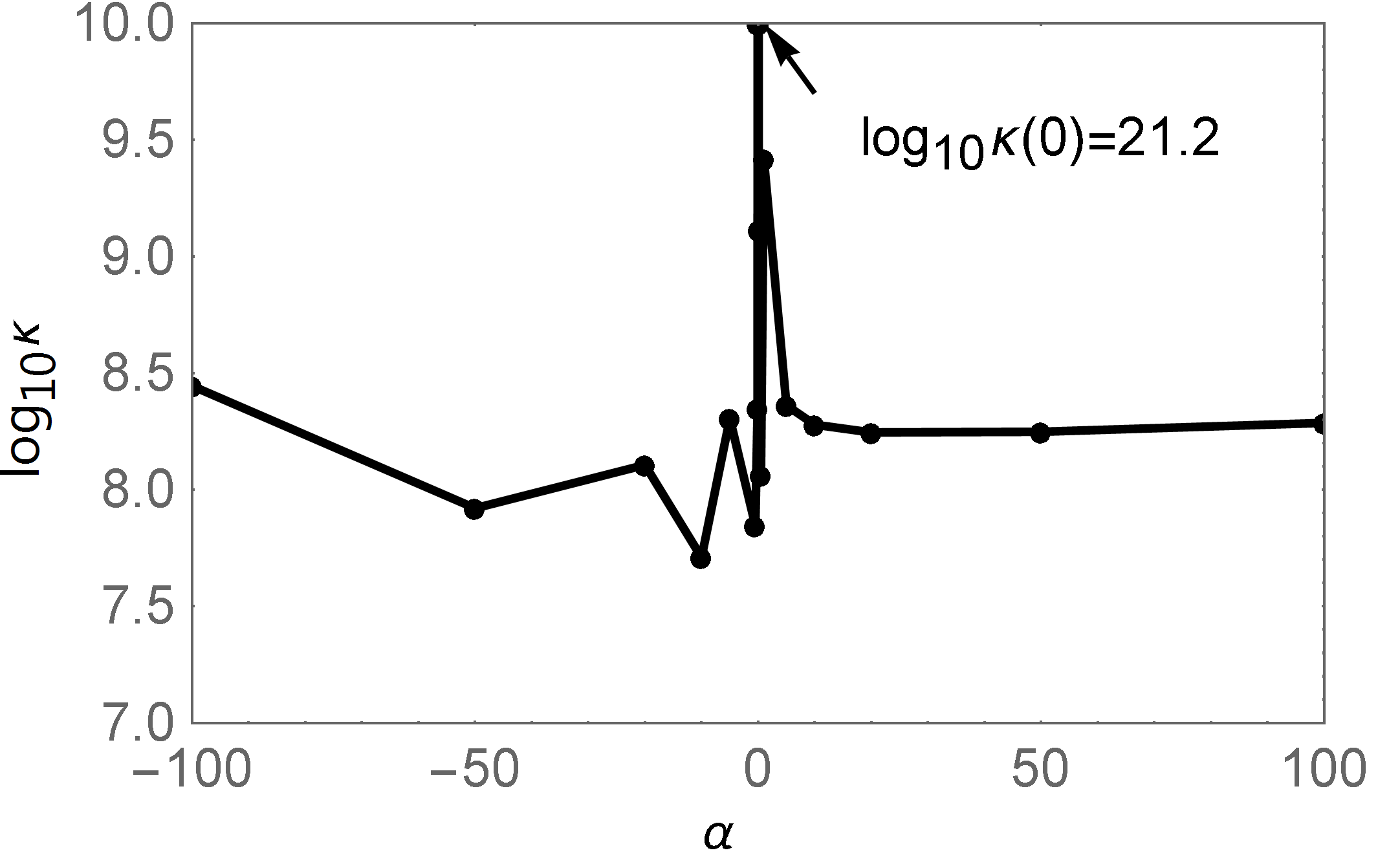}
\caption{Condition number as a function of gauge fixing.}
\label{fig:condNumberPEa}
\end{subfigure}
\begin{subfigure}{\columnwidth}
\includegraphics[width=0.75\columnwidth]{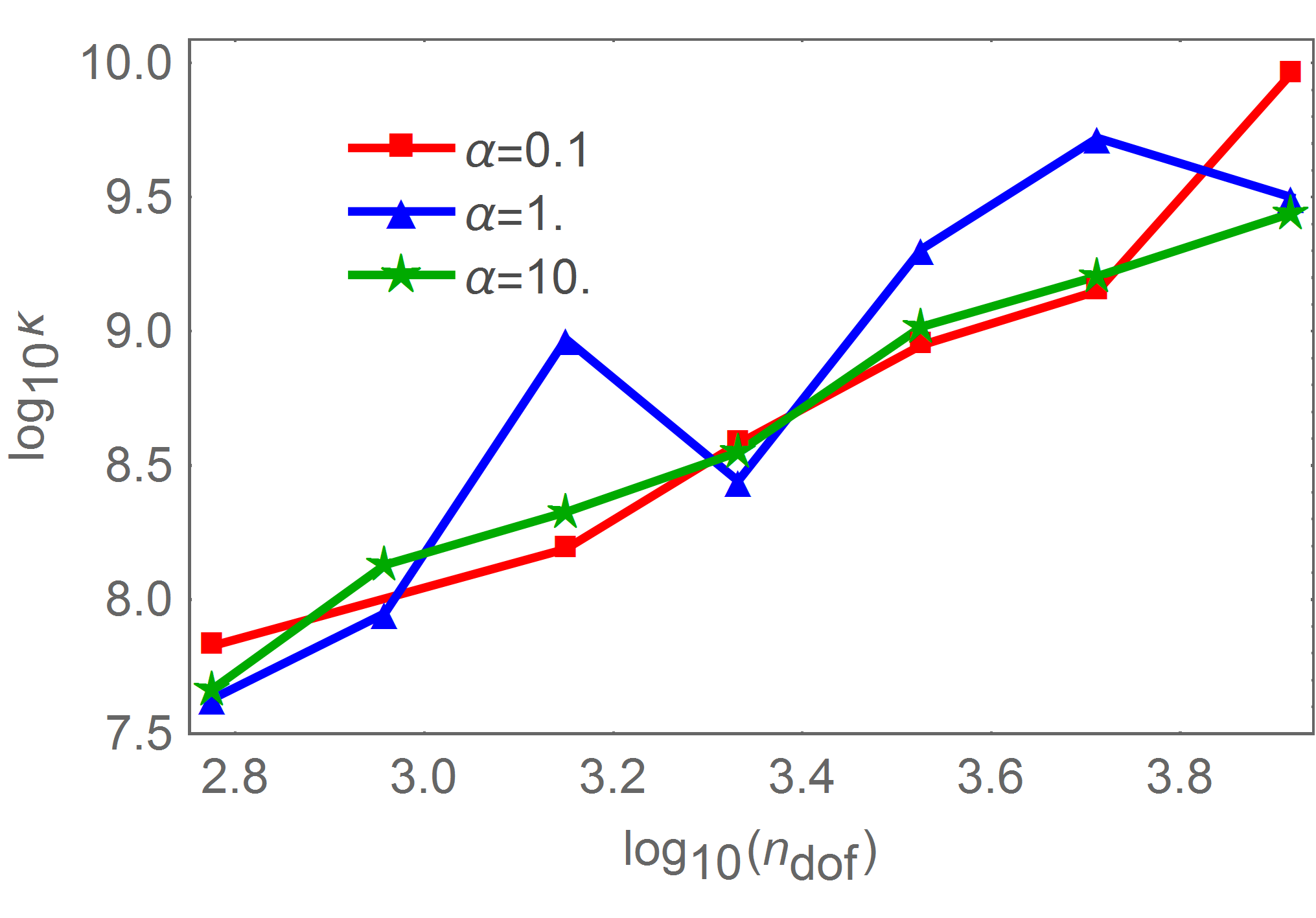}
\caption{Condition number as a function of mesh resolution.}
\label{fig:condNumberPEm}
\end{subfigure}
\caption{Condition number, $\kappa$ for the eigenvalue corresponding to the fundamental TM mode of the cavity shown in fig.~\ref{fig:RidgedWaveguide} with perfect electric boundary.}
\label{fig:condNumbPE} 
\end{figure}

There are two particular aspects to robustness that we consider here: numerical conditioning and the question of spurious modes. A rigorous theoretical analysis is beyond the scope of this paper but we have investigated these issues experimentally. For the eigenmode analysis, we use the definition of condition number for a quadratic eigenvalue given in ref. \onlinecite{TisseurQEP}. Figures \ref{fig:condNumbPM} and \ref{fig:condNumbPE} plot the condition number for the quadratic eigenvalue of the notched pillbox with a perfect magnetic and electric boundary condition, respectively. Plotted as a function of $\alpha$, the need for the gauge fixing coefficient is clear from the singularity in the condition number as $\alpha \rightarrow 0$. There is a similar singularity in conditioning at $\alpha = 0$ for the linear system in the driven problem.

The condition number is reasonable for the perfect magnetic boundary condition and there is no strong variation with $\alpha$ or $h_\mathrm{mesh}$. The same cannot be said for the impedance or (in the limit of large conductance) perfect electric boundary condition. The addition of the surface integral to impose the perfect electric boundary negatively impacts the condition number. We have found the condition number for a given eigenvalue to scale linearly with the conductance, $Y = \frac{1}{Z}$ and as O($h_\mathrm{mesh}^{-2}$) when the impedance boundary is applied. The condition number can be mitigated to some extent by refining the mesh on the boundary while maintaining constant mesh in the interior, but future work will focus on resolving this issue more efficiently by modifying the surface integral or its implementation. 

\begin{figure}
\begin{subfigure}{\columnwidth}
\includegraphics[width=\columnwidth]{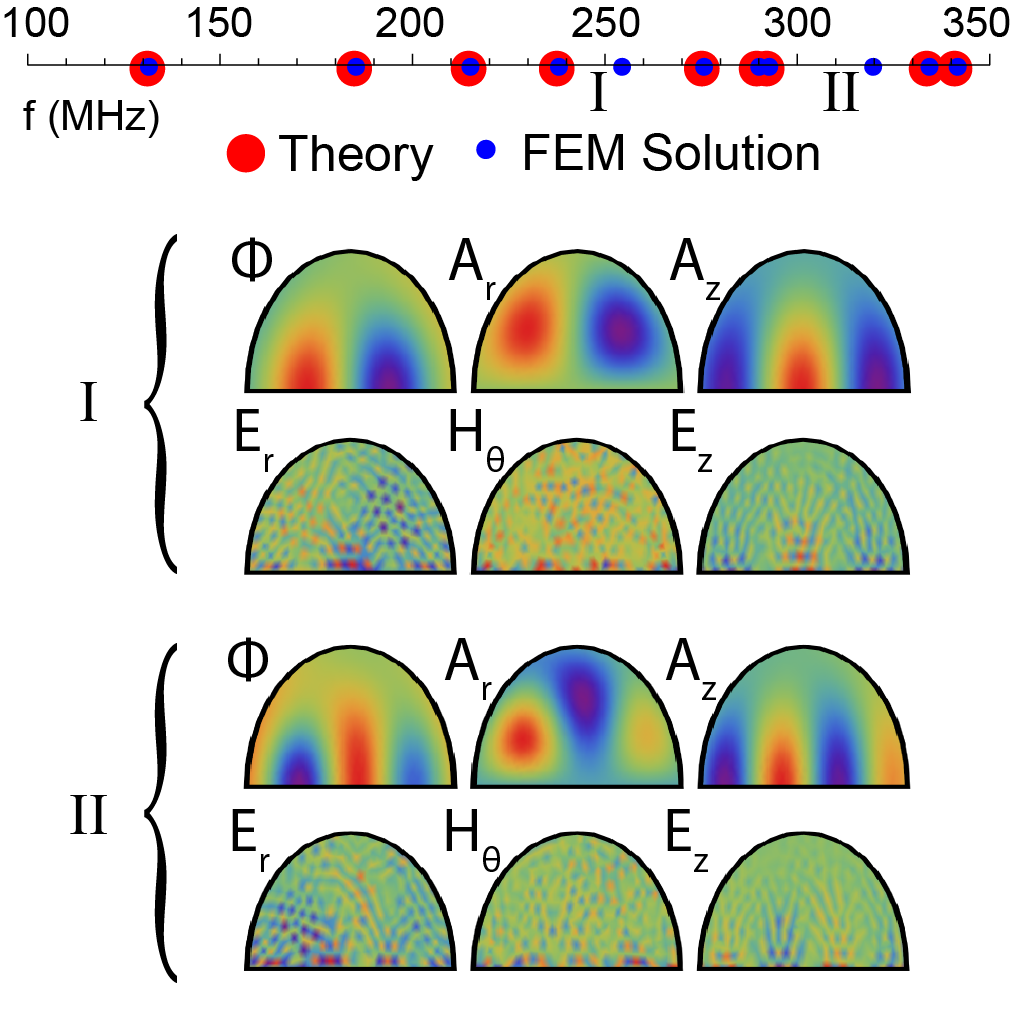}
\caption{Eigenmode}
\label{fig:sphereSpectrum}
\end{subfigure}
\begin{subfigure}{\columnwidth}
\includegraphics[width=\columnwidth]{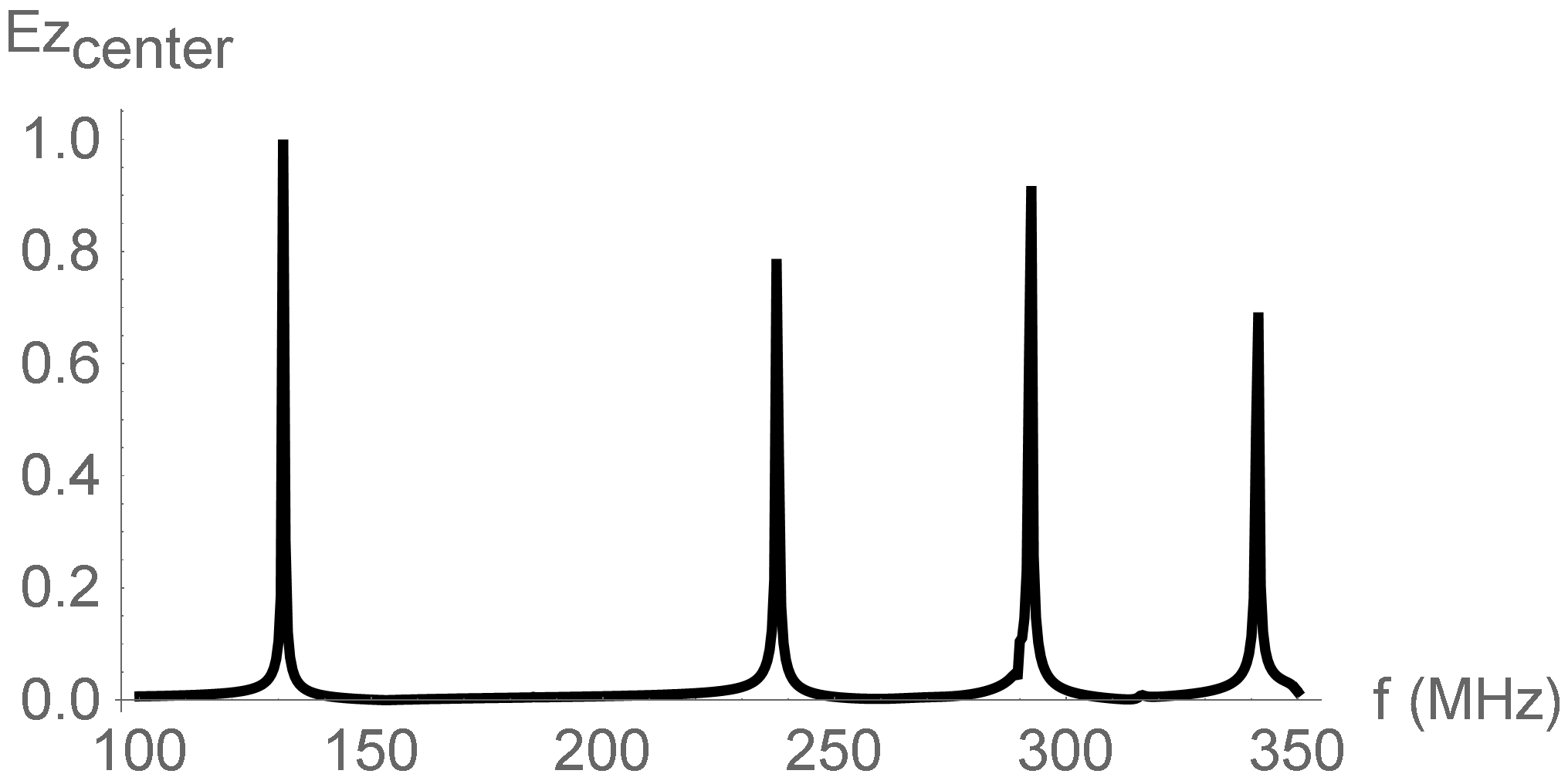}
\caption{Driven: Relative amplitude of $E_z$ at the origin with an applied current, $J_z$, on axis.}
\label{fig:sphereDriven}
\end{subfigure}
\caption{Spectrum showing the first several modes of a spherical cavity with 1m radius as computed through the eigenmode analysis (a) and by scanning the frequency of a driving current and observing the field amplitude (b). The driven spectrum was excited using a current in the $\hat{z}$ direction thus not all resonant modes are reflected in the spectrum. There are two modes in the eigenmode spectrum which are pure gauge modes - valid solutions for the four-potential resulting in vanishing fields. These modes are not excited in the driven problem, as can be seen by the lack of a peak at their respective frequencies.}
\label{fig:sphereZeroSols}
\end{figure}

Up to this point, we have focused on specific modes to demonstrate the convergence and stability of the Lagrangian formulation. It is equally important to ensure that in addition to obtaining correct modes, the solved spectrum is free of unphysical modes. Figure ~\ref{fig:sphereSpectrum} plots the spectrum for the first several solved monopole modes of the spherical cavity. Comparing with the theoretically expected modes, we note the presence of two unexpected modes. These are not spurious modes in the conventional sense, however, but rather are pure gauge modes. As can be seen from the mode profiles, these are valid solutions for the four potential which result in zero field (to within numerical noise). Unlike spurious modes in the nodal curl-curl formulation, these modes converge as the mesh is refined in a similar manner to the expected modes. Most importantly, the number of these pure gauge modes in a given frequency interval does not increase with problem size. In the driven problem, these modes are not excited by sources, as can be seen in fig.~\ref{fig:sphereDriven} (note, only modes with Az on axis are excited so not all resonant modes are present in spectrum). 

\section{Conclusion}
Concluding, we have demonstrated a new finite element formulation to solve time harmonic electromagnetic fields. By encoding the physics of electromagnetism in a different mathematical formulation, the Lagrangian formulation does not suffer from the challenges inherent to the conventional curl-curl equation for $\vec{E}$. In contrast to the curl-curl equation, where $\vec{J}$ is the only driving term, our formulation completely accounts for both $\vec{J}$ and $\rho$. Both Gauss' law and Ampere's law are satisfied, not just globally but over individual elements, in contrast to the commonly employed  N\'ed\'elec edge elements. This is of importance in the analysis of beam driven radiation sources, for example, where the contribution to the fields from the space charge, $\rho$, can be significant even at high frequency. 

We show through both theory and experimental results that the nodal elements are the correct basis choice for our 4D formulation. Indeed, our implementation demonstrates that the four-potential formulation easily handles field singularities and discontinuities unlike nodal element curl-curl implementations. We have benchmarked a proof of concept implementation against COMSOL, a state of the art edge element solver, showing that comparable performance can be obtained. Currently, our surface integral for imposing an impedance or perfect electric boundary condition produces accurate results for problem sizes up to around $n_{DOF}=10^5$. However, the scaling of the condition number with mesh size and conductance needs to be addressed. Finally, we have demonstrated that this approach is not susceptible to spurious modes though pure gauge modes with zero fields do appear in the eigenmode spectrum. 

The Lagrangian formulation provides unique opportunities for the numerical analysis of electromagnetic fields. While here we present some initial results confirming the accuracy, flexibility and robustness of this idea, we believe there is much yet to explore, particularly in the time domain. From a practical point of view, the adoption of the four-potential also offers a straightforward solution for those interested in a nodal field solver. This is not only beneficial in terms of the computational efficiency and simplicity of nodal elements, but given the widespread use of nodal elements in fields from structural mechanics to fluid dynamics, allows for a common framework for multi-physics problems.

\begin{acknowledgments}
This project was funded by U.S. Department of Energy under Contract No. DE-AC02-76SF00515.
\end{acknowledgments}

\bibliographystyle{apsrev4-1} 
\bibliography{References}

\begin{thebibliography}{49}%
\makeatletter
\providecommand \@ifxundefined [1]{%
 \@ifx{#1\undefined}
}%
\providecommand \@ifnum [1]{%
 \ifnum #1\expandafter \@firstoftwo
 \else \expandafter \@secondoftwo
 \fi
}%
\providecommand \@ifx [1]{%
 \ifx #1\expandafter \@firstoftwo
 \else \expandafter \@secondoftwo
 \fi
}%
\providecommand \natexlab [1]{#1}%
\providecommand \enquote  [1]{``#1''}%
\providecommand \bibnamefont  [1]{#1}%
\providecommand \bibfnamefont [1]{#1}%
\providecommand \citenamefont [1]{#1}%
\providecommand \href@noop [0]{\@secondoftwo}%
\providecommand \href [0]{\begingroup \@sanitize@url \@href}%
\providecommand \@href[1]{\@@startlink{#1}\@@href}%
\providecommand \@@href[1]{\endgroup#1\@@endlink}%
\providecommand \@sanitize@url [0]{\catcode `\\12\catcode `\$12\catcode
  `\&12\catcode `\#12\catcode `\^12\catcode `\_12\catcode `\%12\relax}%
\providecommand \@@startlink[1]{}%
\providecommand \@@endlink[0]{}%
\providecommand \url  [0]{\begingroup\@sanitize@url \@url }%
\providecommand \@url [1]{\endgroup\@href {#1}{\urlprefix }}%
\providecommand \urlprefix  [0]{URL }%
\providecommand \Eprint [0]{\href }%
\providecommand \doibase [0]{http://dx.doi.org/}%
\providecommand \selectlanguage [0]{\@gobble}%
\providecommand \bibinfo  [0]{\@secondoftwo}%
\providecommand \bibfield  [0]{\@secondoftwo}%
\providecommand \translation [1]{[#1]}%
\providecommand \BibitemOpen [0]{}%
\providecommand \bibitemStop [0]{}%
\providecommand \bibitemNoStop [0]{.\EOS\space}%
\providecommand \EOS [0]{\spacefactor3000\relax}%
\providecommand \BibitemShut  [1]{\csname bibitem#1\endcsname}%
\let\auto@bib@innerbib\@empty
%</preamble>
\bibitem [{\citenamefont {Jin}(2015)}]{FEMtextbook}%
  \BibitemOpen
  \bibfield  {author} {\bibinfo {author} {\bibfnamefont {J.}~\bibnamefont
  {Jin}},\ }\href@noop {} {\emph {\bibinfo {title} {The Finite Element Method
  in Electromagnetics}}},\ Wiley - IEEE\ (\bibinfo  {publisher} {Wiley},\
  \bibinfo {year} {2015})\BibitemShut {NoStop}%
\bibitem [{\citenamefont {Albanese}\ and\ \citenamefont
  {Rubinacci}(1997)}]{ARreview}%
  \BibitemOpen
  \bibfield  {author} {\bibinfo {author} {\bibfnamefont {R.}~\bibnamefont
  {Albanese}}\ and\ \bibinfo {author} {\bibfnamefont {G.}~\bibnamefont
  {Rubinacci}}\ }(\bibinfo  {publisher} {Elsevier},\ \bibinfo {year} {1997})\
  pp.\ \bibinfo {pages} {1 -- 86}\BibitemShut {NoStop}%
\bibitem [{\citenamefont {{Cortes Garcia}}\ \emph {et~al.}(2018)\citenamefont
  {{Cortes Garcia}}, \citenamefont {{Sch{\"o}ps}}, \citenamefont {{De
  Gersem}},\ and\ \citenamefont {{Baumanns}}}]{ICGreview}%
  \BibitemOpen
  \bibfield  {author} {\bibinfo {author} {\bibfnamefont {I.}~\bibnamefont
  {{Cortes Garcia}}}, \bibinfo {author} {\bibfnamefont {S.}~\bibnamefont
  {{Sch{\"o}ps}}}, \bibinfo {author} {\bibfnamefont {H.}~\bibnamefont {{De
  Gersem}}}, \ and\ \bibinfo {author} {\bibfnamefont {S.}~\bibnamefont
  {{Baumanns}}},\ }\href@noop {} {\bibfield  {journal} {\bibinfo  {journal}
  {arXiv e-prints}\ ,\ \bibinfo {eid} {arXiv:1802.06673}} (\bibinfo {year}
  {2018})},\ \Eprint {http://arxiv.org/abs/1802.06673} {arXiv:1802.06673
  [math.NA]} \BibitemShut {NoStop}%
\bibitem [{\citenamefont {Marder}(1987)}]{marder}%
  \BibitemOpen
  \bibfield  {author} {\bibinfo {author} {\bibfnamefont {B.}~\bibnamefont
  {Marder}},\ }\href {\doibase https://doi.org/10.1016/0021-9991(87)90043-X}
  {\bibfield  {journal} {\bibinfo  {journal} {Journal of Computational
  Physics}\ }\textbf {\bibinfo {volume} {68}},\ \bibinfo {pages} {48 }
  (\bibinfo {year} {1987})}\BibitemShut {NoStop}%
\bibitem [{\citenamefont {Langdon}(1992)}]{LangdonGaussLaw}%
  \BibitemOpen
  \bibfield  {author} {\bibinfo {author} {\bibfnamefont {A.~B.}\ \bibnamefont
  {Langdon}},\ }\href {\doibase https://doi.org/10.1016/0010-4655(92)90105-8}
  {\bibfield  {journal} {\bibinfo  {journal} {Computer Physics Communications}\
  }\textbf {\bibinfo {volume} {70}},\ \bibinfo {pages} {447 } (\bibinfo {year}
  {1992})}\BibitemShut {NoStop}%
\bibitem [{\citenamefont {CIARLET}\ and\ \citenamefont
  {LABRUNIE}(2009)}]{CiarletGaussLaw}%
  \BibitemOpen
  \bibfield  {author} {\bibinfo {author} {\bibfnamefont {P.}~\bibnamefont
  {CIARLET}}\ and\ \bibinfo {author} {\bibfnamefont {S.}~\bibnamefont
  {LABRUNIE}},\ }\href {\doibase 10.1142/S0218202509004017} {\bibfield
  {journal} {\bibinfo  {journal} {Mathematical Models and Methods in Applied
  Sciences}\ }\textbf {\bibinfo {volume} {19}},\ \bibinfo {pages} {1959}
  (\bibinfo {year} {2009})}\BibitemShut {NoStop}%
\bibitem [{\citenamefont {BARTHELMÉ}\ \emph {et~al.}(2007)\citenamefont
  {BARTHELMÉ}, \citenamefont {CIARLET},\ and\ \citenamefont
  {SONNENDRÜCKER}}]{GeneralizedFormulationForVlasovMaxwell}%
  \BibitemOpen
  \bibfield  {author} {\bibinfo {author} {\bibfnamefont {R.}~\bibnamefont
  {BARTHELMÉ}}, \bibinfo {author} {\bibfnamefont {P.}~\bibnamefont {CIARLET}},
  \ and\ \bibinfo {author} {\bibfnamefont {E.}~\bibnamefont {SONNENDRÜCKER}},\
  }\href {\doibase 10.1142/S0218202507002066} {\bibfield  {journal} {\bibinfo
  {journal} {Mathematical Models and Methods in Applied Sciences}\ }\textbf
  {\bibinfo {volume} {17}},\ \bibinfo {pages} {657} (\bibinfo {year}
  {2007})}\BibitemShut {NoStop}%
\bibitem [{\citenamefont {Ciarlet}\ \emph {et~al.}(2014)\citenamefont
  {Ciarlet}, \citenamefont {Wu},\ and\ \citenamefont
  {Zou}}]{EdgeElementCiarlet}%
  \BibitemOpen
  \bibfield  {author} {\bibinfo {author} {\bibfnamefont {P.}~\bibnamefont
  {Ciarlet}}, \bibinfo {author} {\bibfnamefont {H.}~\bibnamefont {Wu}}, \ and\
  \bibinfo {author} {\bibfnamefont {J.}~\bibnamefont {Zou}},\ }\href {\doibase
  10.1137/120899856} {\bibfield  {journal} {\bibinfo  {journal} {SIAM Journal
  on Numerical Analysis}\ }\textbf {\bibinfo {volume} {52}},\ \bibinfo {pages}
  {779} (\bibinfo {year} {2014})}\BibitemShut {NoStop}%
\bibitem [{\citenamefont {Li}\ \emph {et~al.}(2016)\citenamefont {Li},
  \citenamefont {Sun}, \citenamefont {Dai},\ and\ \citenamefont
  {Chew}}]{LorenzGaugedMixed}%
  \BibitemOpen
  \bibfield  {author} {\bibinfo {author} {\bibfnamefont {Y.}~\bibnamefont
  {Li}}, \bibinfo {author} {\bibfnamefont {S.}~\bibnamefont {Sun}}, \bibinfo
  {author} {\bibfnamefont {Q.~I.}\ \bibnamefont {Dai}}, \ and\ \bibinfo
  {author} {\bibfnamefont {W.~C.}\ \bibnamefont {Chew}},\ }\href {\doibase
  10.1109/TAP.2016.2593748} {\bibfield  {journal} {\bibinfo  {journal} {IEEE
  Transactions on Antennas and Propagation}\ }\textbf {\bibinfo {volume}
  {64}},\ \bibinfo {pages} {4355} (\bibinfo {year} {2016})}\BibitemShut
  {NoStop}%
\bibitem [{\citenamefont {{Hiptmair}}\ \emph {et~al.}(2008)\citenamefont
  {{Hiptmair}}, \citenamefont {{Kramer}},\ and\ \citenamefont
  {{Ostrowski}}}]{hiptmair}%
  \BibitemOpen
  \bibfield  {author} {\bibinfo {author} {\bibfnamefont {R.}~\bibnamefont
  {{Hiptmair}}}, \bibinfo {author} {\bibfnamefont {F.}~\bibnamefont
  {{Kramer}}}, \ and\ \bibinfo {author} {\bibfnamefont {J.}~\bibnamefont
  {{Ostrowski}}},\ }\href {\doibase 10.1109/TMAG.2007.915991} {\bibfield
  {journal} {\bibinfo  {journal} {IEEE Transactions on Magnetics}\ }\textbf
  {\bibinfo {volume} {44}},\ \bibinfo {pages} {682} (\bibinfo {year}
  {2008})}\BibitemShut {NoStop}%
\bibitem [{\citenamefont {Stoker}(1989)}]{diffGeoTextbook}%
  \BibitemOpen
  \bibfield  {author} {\bibinfo {author} {\bibfnamefont {J.}~\bibnamefont
  {Stoker}},\ }\href {https://books.google.com/books?id=5Qk7WwXePuQC} {\emph
  {\bibinfo {title} {Differential Geometry}}},\ Pure and applied mathematics\
  (\bibinfo  {publisher} {Wiley},\ \bibinfo {year} {1989})\BibitemShut
  {NoStop}%
\bibitem [{\citenamefont {{Hara}}\ \emph {et~al.}(1983)\citenamefont {{Hara}},
  \citenamefont {{Wada}}, \citenamefont {{Fukasawa}},\ and\ \citenamefont
  {{Kikuchi}}}]{penaltymethod1983}%
  \BibitemOpen
  \bibfield  {author} {\bibinfo {author} {\bibfnamefont {M.}~\bibnamefont
  {{Hara}}}, \bibinfo {author} {\bibfnamefont {T.}~\bibnamefont {{Wada}}},
  \bibinfo {author} {\bibfnamefont {T.}~\bibnamefont {{Fukasawa}}}, \ and\
  \bibinfo {author} {\bibfnamefont {F.}~\bibnamefont {{Kikuchi}}},\ }\href
  {\doibase 10.1109/TNS.1983.4336751} {\bibfield  {journal} {\bibinfo
  {journal} {IEEE Transactions on Nuclear Science}\ }\textbf {\bibinfo {volume}
  {30}},\ \bibinfo {pages} {3639} (\bibinfo {year} {1983})}\BibitemShut
  {NoStop}%
\bibitem [{\citenamefont {{Rahman}}\ and\ \citenamefont
  {{Davies}}(1984)}]{penaltymethod}%
  \BibitemOpen
  \bibfield  {author} {\bibinfo {author} {\bibfnamefont {B.~M.~A.}\
  \bibnamefont {{Rahman}}}\ and\ \bibinfo {author} {\bibfnamefont {J.~B.}\
  \bibnamefont {{Davies}}},\ }\href {\doibase 10.1109/TMTT.1984.1132789}
  {\bibfield  {journal} {\bibinfo  {journal} {IEEE Transactions on Microwave
  Theory and Techniques}\ }\textbf {\bibinfo {volume} {32}},\ \bibinfo {pages}
  {922} (\bibinfo {year} {1984})}\BibitemShut {NoStop}%
\bibitem [{\citenamefont {Demkowicz}(2017)}]{FEMmaxwellChapter}%
  \BibitemOpen
  \bibfield  {author} {\bibinfo {author} {\bibfnamefont {L.}~\bibnamefont
  {Demkowicz}},\ }\enquote {\bibinfo {title} {Finite element methods for
  maxwell's equations},}\ in\ \href@noop {} {\emph {\bibinfo {booktitle}
  {Encyclopedia of Computational Mechanics Second Edition}}}\ (\bibinfo
  {publisher} {American Cancer Society},\ \bibinfo {year} {2017})\ pp.\
  \bibinfo {pages} {1--20}\BibitemShut {NoStop}%
\bibitem [{\citenamefont {Adam}\ \emph {et~al.}(1997)\citenamefont {Adam},
  \citenamefont {Arbenz},\ and\ \citenamefont {Geus}}]{ETHEigenSolver}%
  \BibitemOpen
  \bibfield  {author} {\bibinfo {author} {\bibfnamefont {S.}~\bibnamefont
  {Adam}}, \bibinfo {author} {\bibfnamefont {P.}~\bibnamefont {Arbenz}}, \ and\
  \bibinfo {author} {\bibfnamefont {R.}~\bibnamefont {Geus}},\ }\href@noop {}
  {\bibfield  {journal} {\bibinfo  {journal} {Technische Berichte/ETH
  Z{\"u}rich, Departement Informatik}\ }\textbf {\bibinfo {volume} {275}}
  (\bibinfo {year} {1997})}\BibitemShut {NoStop}%
\bibitem [{\citenamefont {Costabel}\ and\ \citenamefont
  {Dauge}(2003)}]{CDlong}%
  \BibitemOpen
  \bibfield  {author} {\bibinfo {author} {\bibfnamefont {M.}~\bibnamefont
  {Costabel}}\ and\ \bibinfo {author} {\bibfnamefont {M.}~\bibnamefont
  {Dauge}}\ }(\bibinfo  {publisher} {Springer},\ \bibinfo {year} {2003})\ pp.\
  \bibinfo {pages} {125--161}\BibitemShut {NoStop}%
\bibitem [{\citenamefont {Costabel}\ and\ \citenamefont
  {Dauge}(2000)}]{CDarticle}%
  \BibitemOpen
  \bibfield  {author} {\bibinfo {author} {\bibfnamefont {M.}~\bibnamefont
  {Costabel}}\ and\ \bibinfo {author} {\bibfnamefont {M.}~\bibnamefont
  {Dauge}},\ }\href@noop {} {\bibfield  {journal} {\bibinfo  {journal} {Archive
  for Rational Mechanics and Analysis}\ }\textbf {\bibinfo {volume} {151}},\
  \bibinfo {pages} {221} (\bibinfo {year} {2000})}\BibitemShut {NoStop}%
\bibitem [{\citenamefont {Assous}\ \emph {et~al.}(1999)\citenamefont {Assous},
  \citenamefont {Ciarlet~Jr.}, \citenamefont {Raviart},\ and\ \citenamefont
  {Sonnendrücker}}]{AssousSingularDecomp}%
  \BibitemOpen
  \bibfield  {author} {\bibinfo {author} {\bibfnamefont {F.}~\bibnamefont
  {Assous}}, \bibinfo {author} {\bibfnamefont {P.}~\bibnamefont {Ciarlet~Jr.}},
  \bibinfo {author} {\bibfnamefont {P.-A.}\ \bibnamefont {Raviart}}, \ and\
  \bibinfo {author} {\bibfnamefont {E.}~\bibnamefont {Sonnendrücker}},\ }\href
  {\doibase 10.1002/(SICI)1099-1476(199904)22:6<485::AID-MMA46>3.0.CO;2-E}
  {\bibfield  {journal} {\bibinfo  {journal} {Mathematical Methods in the
  Applied Sciences}\ }\textbf {\bibinfo {volume} {22}},\ \bibinfo {pages} {485}
  (\bibinfo {year} {1999})}\BibitemShut {NoStop}%
\bibitem [{\citenamefont {Dhia}\ \emph {et~al.}(1999)\citenamefont {Dhia},
  \citenamefont {Hazard},\ and\ \citenamefont {Lohrengel}}]{SingFieldMethod}%
  \BibitemOpen
  \bibfield  {author} {\bibinfo {author} {\bibfnamefont {A.}~\bibnamefont
  {Dhia}}, \bibinfo {author} {\bibfnamefont {C.}~\bibnamefont {Hazard}}, \ and\
  \bibinfo {author} {\bibfnamefont {S.}~\bibnamefont {Lohrengel}},\ }\href@noop
  {} {\bibfield  {journal} {\bibinfo  {journal} {SIAM Journal on Applied
  Mathematics}\ }\textbf {\bibinfo {volume} {59}},\ \bibinfo {pages} {2028}
  (\bibinfo {year} {1999})}\BibitemShut {NoStop}%
\bibitem [{\citenamefont {Assous}\ \emph {et~al.}(1998)\citenamefont {Assous},
  \citenamefont {Ciarlet},\ and\ \citenamefont
  {Sonnendr{\"u}cker}}]{AssousSingular}%
  \BibitemOpen
  \bibfield  {author} {\bibinfo {author} {\bibfnamefont {F.}~\bibnamefont
  {Assous}}, \bibinfo {author} {\bibfnamefont {P.}~\bibnamefont {Ciarlet}}, \
  and\ \bibinfo {author} {\bibfnamefont {E.}~\bibnamefont
  {Sonnendr{\"u}cker}},\ }\href
  {https://hal-ensta.archives-ouvertes.fr/hal-01010426} {\bibfield  {journal}
  {\bibinfo  {journal} {{Modelisation Math{\'e}matique et Analyse
  Num{\'e}rique}}\ }\textbf {\bibinfo {volume} {32}},\ \bibinfo {pages} {359}
  (\bibinfo {year} {1998})}\BibitemShut {NoStop}%
\bibitem [{\citenamefont {Duan}\ \emph {et~al.}(2014)\citenamefont {Duan},
  \citenamefont {Tan}, \citenamefont {Yang},\ and\ \citenamefont
  {You}}]{bubbleElements}%
  \BibitemOpen
  \bibfield  {author} {\bibinfo {author} {\bibfnamefont {H.-Y.}\ \bibnamefont
  {Duan}}, \bibinfo {author} {\bibfnamefont {R.~C.}\ \bibnamefont {Tan}},
  \bibinfo {author} {\bibfnamefont {S.-Y.}\ \bibnamefont {Yang}}, \ and\
  \bibinfo {author} {\bibfnamefont {C.-S.}\ \bibnamefont {You}},\ }\href
  {\doibase https://doi.org/10.1016/j.jcp.2014.02.044} {\bibfield  {journal}
  {\bibinfo  {journal} {Journal of Computational Physics}\ }\textbf {\bibinfo
  {volume} {268}},\ \bibinfo {pages} {63 } (\bibinfo {year}
  {2014})}\BibitemShut {NoStop}%
\bibitem [{\citenamefont {Costabel}\ and\ \citenamefont
  {Dauge}(2002)}]{CDweighted}%
  \BibitemOpen
  \bibfield  {author} {\bibinfo {author} {\bibfnamefont {M.}~\bibnamefont
  {Costabel}}\ and\ \bibinfo {author} {\bibfnamefont {M.}~\bibnamefont
  {Dauge}},\ }\href@noop {} {\bibfield  {journal} {\bibinfo  {journal}
  {Numerische Mathematik}\ }\textbf {\bibinfo {volume} {93}},\ \bibinfo {pages}
  {239} (\bibinfo {year} {2002})}\BibitemShut {NoStop}%
\bibitem [{\citenamefont {{Ciarlet Jr., Patrick}}\ \emph
  {et~al.}(2010)\citenamefont {{Ciarlet Jr., Patrick}}, \citenamefont
  {{Lef\`evre, Fran\c{c}ois}}, \citenamefont {{Lohrengel, St\'ephanie}},\ and\
  \citenamefont {{Nicaise, Serge}}}]{CiarletWeighted}%
  \BibitemOpen
  \bibfield  {author} {\bibinfo {author} {\bibnamefont {{Ciarlet Jr.,
  Patrick}}}, \bibinfo {author} {\bibnamefont {{Lef\`evre, Fran\c{c}ois}}},
  \bibinfo {author} {\bibnamefont {{Lohrengel, St\'ephanie}}}, \ and\ \bibinfo
  {author} {\bibnamefont {{Nicaise, Serge}}},\ }\href {\doibase
  10.1051/m2an/2009041} {\bibfield  {journal} {\bibinfo  {journal} {ESAIM:
  M2AN}\ }\textbf {\bibinfo {volume} {44}},\ \bibinfo {pages} {75} (\bibinfo
  {year} {2010})}\BibitemShut {NoStop}%
\bibitem [{\citenamefont {Otin}(2010)}]{otin2010regularized}%
  \BibitemOpen
  \bibfield  {author} {\bibinfo {author} {\bibfnamefont {R.}~\bibnamefont
  {Otin}},\ }\href@noop {} {\bibfield  {journal} {\bibinfo  {journal}
  {Electromagnetics}\ }\textbf {\bibinfo {volume} {30}},\ \bibinfo {pages}
  {190} (\bibinfo {year} {2010})}\BibitemShut {NoStop}%
\bibitem [{\citenamefont {Whitney}(1957)}]{WhitneyBook}%
  \BibitemOpen
  \bibfield  {author} {\bibinfo {author} {\bibfnamefont {H.}~\bibnamefont
  {Whitney}},\ }\href@noop {} {\emph {\bibinfo {title} {Geometric Integration
  Theory}}}\ (\bibinfo  {publisher} {Princeton University Press},\ \bibinfo
  {address} {Princeton},\ \bibinfo {year} {1957})\BibitemShut {NoStop}%
\bibitem [{\citenamefont {Nedelec}(1980)}]{NedelecElements}%
  \BibitemOpen
  \bibfield  {author} {\bibinfo {author} {\bibfnamefont {J.~C.}\ \bibnamefont
  {Nedelec}},\ }\href {\doibase 10.1007/BF01396415} {\bibfield  {journal}
  {\bibinfo  {journal} {Numer. Math.}\ }\textbf {\bibinfo {volume} {35}},\
  \bibinfo {pages} {315} (\bibinfo {year} {1980})}\BibitemShut {NoStop}%
\bibitem [{\citenamefont {Bossavit}(1988)}]{WhitneyForms}%
  \BibitemOpen
  \bibfield  {author} {\bibinfo {author} {\bibfnamefont {A.}~\bibnamefont
  {Bossavit}},\ }\href {\doibase 10.1049/ip-a-1.1988.0077} {\bibfield
  {journal} {\bibinfo  {journal} {IEE Proceedings A - Physical Science,
  Measurement and Instrumentation, Management and Education - Reviews}\
  }\textbf {\bibinfo {volume} {135}},\ \bibinfo {pages} {493} (\bibinfo {year}
  {1988})}\BibitemShut {NoStop}%
\bibitem [{\citenamefont {Bossavit}\ and\ \citenamefont
  {Mayergoyz}(1998)}]{BossavitTextbook}%
  \BibitemOpen
  \bibfield  {author} {\bibinfo {author} {\bibfnamefont {A.}~\bibnamefont
  {Bossavit}}\ and\ \bibinfo {author} {\bibfnamefont {I.}~\bibnamefont
  {Mayergoyz}},\ }\href@noop {} {\emph {\bibinfo {title} {Computational
  Electromagnetism: Variational Formulations, Complementarity, Edge
  Elements}}},\ Electromagnetism\ (\bibinfo  {publisher} {Elsevier Science},\
  \bibinfo {year} {1998})\BibitemShut {NoStop}%
\bibitem [{\citenamefont {Albanese}\ and\ \citenamefont
  {Rubinacci}(1988)}]{AlbaneseRubinacciTCT}%
  \BibitemOpen
  \bibfield  {author} {\bibinfo {author} {\bibfnamefont {R.}~\bibnamefont
  {Albanese}}\ and\ \bibinfo {author} {\bibfnamefont {G.}~\bibnamefont
  {Rubinacci}},\ }\href {\doibase 10.1109/20.43865} {\bibfield  {journal}
  {\bibinfo  {journal} {IEEE Transactions on Magnetics}\ }\textbf {\bibinfo
  {volume} {24}},\ \bibinfo {pages} {98} (\bibinfo {year} {1988})}\BibitemShut
  {NoStop}%
\bibitem [{\citenamefont {Trapp}\ \emph {et~al.}(2002)\citenamefont {Trapp},
  \citenamefont {Munteanu}, \citenamefont {Schuhmann}, \citenamefont
  {Weiland},\ and\ \citenamefont {Ioan}}]{TrappEigenvalueTCT}%
  \BibitemOpen
  \bibfield  {author} {\bibinfo {author} {\bibfnamefont {B.}~\bibnamefont
  {Trapp}}, \bibinfo {author} {\bibfnamefont {H.}~\bibnamefont {Munteanu}},
  \bibinfo {author} {\bibfnamefont {R.}~\bibnamefont {Schuhmann}}, \bibinfo
  {author} {\bibfnamefont {T.}~\bibnamefont {Weiland}}, \ and\ \bibinfo
  {author} {\bibfnamefont {D.}~\bibnamefont {Ioan}},\ }\href {\doibase
  10.1109/20.996118} {\bibfield  {journal} {\bibinfo  {journal} {IEEE
  Transactions on Magnetics}\ }\textbf {\bibinfo {volume} {38}},\ \bibinfo
  {pages} {445} (\bibinfo {year} {2002})}\BibitemShut {NoStop}%
\bibitem [{\citenamefont {Wang}\ \emph {et~al.}(2010)\citenamefont {Wang},
  \citenamefont {Riley},\ and\ \citenamefont {Jin}}]{WangTimeDomainTCT}%
  \BibitemOpen
  \bibfield  {author} {\bibinfo {author} {\bibfnamefont {R.}~\bibnamefont
  {Wang}}, \bibinfo {author} {\bibfnamefont {D.~J.}\ \bibnamefont {Riley}}, \
  and\ \bibinfo {author} {\bibfnamefont {J.}~\bibnamefont {Jin}},\ }\href
  {\doibase 10.1109/TAP.2010.2044348} {\bibfield  {journal} {\bibinfo
  {journal} {IEEE Transactions on Antennas and Propagation}\ }\textbf {\bibinfo
  {volume} {58}},\ \bibinfo {pages} {1590} (\bibinfo {year}
  {2010})}\BibitemShut {NoStop}%
\bibitem [{\citenamefont {Manges}\ and\ \citenamefont
  {Cendes}(1997)}]{magnesTCToverview}%
  \BibitemOpen
  \bibfield  {author} {\bibinfo {author} {\bibfnamefont {J.}~\bibnamefont
  {Manges}}\ and\ \bibinfo {author} {\bibfnamefont {Z.}~\bibnamefont
  {Cendes}},\ }\href {\doibase
  10.1002/(SICI)1097-0207(19970515)40:9<1667::AID-NME133>3.0.CO;2-9} {\bibfield
   {journal} {\bibinfo  {journal} {International Journal for Numerical Methods
  in Engineering}\ }\textbf {\bibinfo {volume} {40}},\ \bibinfo {pages} {1667}
  (\bibinfo {year} {1997})}\BibitemShut {NoStop}%
\bibitem [{\citenamefont {Ticar}\ \emph {et~al.}(2002)\citenamefont {Ticar},
  \citenamefont {Biro},\ and\ \citenamefont {Preis}}]{TicarBiro}%
  \BibitemOpen
  \bibfield  {author} {\bibinfo {author} {\bibfnamefont {I.}~\bibnamefont
  {Ticar}}, \bibinfo {author} {\bibfnamefont {O.}~\bibnamefont {Biro}}, \ and\
  \bibinfo {author} {\bibfnamefont {K.}~\bibnamefont {Preis}},\ }\href
  {\doibase 10.1109/20.996116} {\bibfield  {journal} {\bibinfo  {journal} {IEEE
  Transactions on Magnetics}\ }\textbf {\bibinfo {volume} {38}},\ \bibinfo
  {pages} {437} (\bibinfo {year} {2002})}\BibitemShut {NoStop}%
\bibitem [{\citenamefont {Golias}\ and\ \citenamefont
  {Tsiboukis}(1994)}]{GoliasTreeChoice}%
  \BibitemOpen
  \bibfield  {author} {\bibinfo {author} {\bibfnamefont {N.~A.}\ \bibnamefont
  {Golias}}\ and\ \bibinfo {author} {\bibfnamefont {T.~D.}\ \bibnamefont
  {Tsiboukis}},\ }\href {\doibase 10.1109/20.312537} {\bibfield  {journal}
  {\bibinfo  {journal} {IEEE Transactions on Magnetics}\ }\textbf {\bibinfo
  {volume} {30}},\ \bibinfo {pages} {2877} (\bibinfo {year}
  {1994})}\BibitemShut {NoStop}%
\bibitem [{\citenamefont {Preis}\ \emph {et~al.}(1992)\citenamefont {Preis},
  \citenamefont {Bardi}, \citenamefont {Biro}, \citenamefont {Magele},
  \citenamefont {Vrisk},\ and\ \citenamefont {Richter}}]{PreisNodalEdgeTbad}%
  \BibitemOpen
  \bibfield  {author} {\bibinfo {author} {\bibfnamefont {K.}~\bibnamefont
  {Preis}}, \bibinfo {author} {\bibfnamefont {I.}~\bibnamefont {Bardi}},
  \bibinfo {author} {\bibfnamefont {O.}~\bibnamefont {Biro}}, \bibinfo {author}
  {\bibfnamefont {C.}~\bibnamefont {Magele}}, \bibinfo {author} {\bibfnamefont
  {G.}~\bibnamefont {Vrisk}}, \ and\ \bibinfo {author} {\bibfnamefont {K.~R.}\
  \bibnamefont {Richter}},\ }\href {\doibase 10.1109/20.123863} {\bibfield
  {journal} {\bibinfo  {journal} {IEEE Transactions on Magnetics}\ }\textbf
  {\bibinfo {volume} {28}},\ \bibinfo {pages} {1056} (\bibinfo {year}
  {1992})}\BibitemShut {NoStop}%
\bibitem [{\citenamefont {Ahagon}\ and\ \citenamefont
  {Kameari}(2017)}]{AhagonKameari}%
  \BibitemOpen
  \bibfield  {author} {\bibinfo {author} {\bibfnamefont {A.}~\bibnamefont
  {Ahagon}}\ and\ \bibinfo {author} {\bibfnamefont {A.}~\bibnamefont
  {Kameari}},\ }\href {\doibase 10.1109/TMAG.2017.2658680} {\bibfield
  {journal} {\bibinfo  {journal} {IEEE Transactions on Magnetics}\ }\textbf
  {\bibinfo {volume} {53}},\ \bibinfo {pages} {1} (\bibinfo {year}
  {2017})}\BibitemShut {NoStop}%
\bibitem [{\citenamefont {Schwartz}(2014)}]{QFTSchwartz}%
  \BibitemOpen
  \bibfield  {author} {\bibinfo {author} {\bibfnamefont {M.}~\bibnamefont
  {Schwartz}},\ }\href {https://books.google.ca/books?id=HbdEAgAAQBAJ} {\emph
  {\bibinfo {title} {Quantum Field Theory and the Standard Model}}},\ Quantum
  Field Theory and the Standard Model\ (\bibinfo  {publisher} {Cambridge
  University Press},\ \bibinfo {year} {2014})\ Chap.\ \bibinfo {chapter}
  {8.5.1}\BibitemShut {NoStop}%
\bibitem [{\citenamefont {{Paulsen}}\ \emph {et~al.}(1992)\citenamefont
  {{Paulsen}}, \citenamefont {{Boyse}},\ and\ \citenamefont
  {{Lynch}}}]{BoyseMainPaperOnNodalPotentialFormulation}%
  \BibitemOpen
  \bibfield  {author} {\bibinfo {author} {\bibfnamefont {K.~D.}\ \bibnamefont
  {{Paulsen}}}, \bibinfo {author} {\bibfnamefont {W.~E.}\ \bibnamefont
  {{Boyse}}}, \ and\ \bibinfo {author} {\bibfnamefont {D.~R.}\ \bibnamefont
  {{Lynch}}},\ }\href {\doibase 10.1109/8.182451} {\bibfield  {journal}
  {\bibinfo  {journal} {IEEE Transactions on Antennas and Propagation}\
  }\textbf {\bibinfo {volume} {40}},\ \bibinfo {pages} {1192} (\bibinfo {year}
  {1992})}\BibitemShut {NoStop}%
\bibitem [{\citenamefont {{Boyse}}\ \emph {et~al.}(1992)\citenamefont
  {{Boyse}}, \citenamefont {{Lynch}}, \citenamefont {{Paulsen}},\ and\
  \citenamefont {{Minerbo}}}]{BoyseTheoretical}%
  \BibitemOpen
  \bibfield  {author} {\bibinfo {author} {\bibfnamefont {W.~E.}\ \bibnamefont
  {{Boyse}}}, \bibinfo {author} {\bibfnamefont {D.~R.}\ \bibnamefont
  {{Lynch}}}, \bibinfo {author} {\bibfnamefont {K.~D.}\ \bibnamefont
  {{Paulsen}}}, \ and\ \bibinfo {author} {\bibfnamefont {G.~N.}\ \bibnamefont
  {{Minerbo}}},\ }\href {\doibase 10.1109/8.144598} {\bibfield  {journal}
  {\bibinfo  {journal} {IEEE Transactions on Antennas and Propagation}\
  }\textbf {\bibinfo {volume} {40}},\ \bibinfo {pages} {642} (\bibinfo {year}
  {1992})}\BibitemShut {NoStop}%
\bibitem [{\citenamefont {Jog}\ and\ \citenamefont
  {Nandy}(2014)}]{PotentialNotWorking}%
  \BibitemOpen
  \bibfield  {author} {\bibinfo {author} {\bibfnamefont {C.}~\bibnamefont
  {Jog}}\ and\ \bibinfo {author} {\bibfnamefont {A.}~\bibnamefont {Nandy}},\
  }\href@noop {} {\bibfield  {journal} {\bibinfo  {journal} {Computers and
  Mathematics with Applications}\ }\textbf {\bibinfo {volume} {68}},\ \bibinfo
  {pages} {887 } (\bibinfo {year} {2014})}\BibitemShut {NoStop}%
\bibitem [{\citenamefont {Duan}\ \emph {et~al.}(2018)\citenamefont {Duan},
  \citenamefont {Tan}, \citenamefont {Yang},\ and\ \citenamefont
  {You}}]{H1MixedDuan}%
  \BibitemOpen
  \bibfield  {author} {\bibinfo {author} {\bibfnamefont {H.}~\bibnamefont
  {Duan}}, \bibinfo {author} {\bibfnamefont {R.}~\bibnamefont {Tan}}, \bibinfo
  {author} {\bibfnamefont {S.}~\bibnamefont {Yang}}, \ and\ \bibinfo {author}
  {\bibfnamefont {C.}~\bibnamefont {You}},\ }\href {\doibase
  10.1137/16M1078082} {\bibfield  {journal} {\bibinfo  {journal} {SIAM Journal
  on Scientific Computing}\ }\textbf {\bibinfo {volume} {40}},\ \bibinfo
  {pages} {A224} (\bibinfo {year} {2018})}\BibitemShut {NoStop}%
\bibitem [{\citenamefont {Baumanns}\ \emph {et~al.}(2013)\citenamefont
  {Baumanns}, \citenamefont {Clemens},\ and\ \citenamefont
  {Schöps}}]{FullFITEMBaumanns}%
  \BibitemOpen
  \bibfield  {author} {\bibinfo {author} {\bibfnamefont {S.}~\bibnamefont
  {Baumanns}}, \bibinfo {author} {\bibfnamefont {M.}~\bibnamefont {Clemens}}, \
  and\ \bibinfo {author} {\bibfnamefont {S.}~\bibnamefont {Schöps}},\ }in\
  \href@noop {} {\emph {\bibinfo {booktitle} {2013 International Symposium on
  Electromagnetic Theory}}}\ (\bibinfo {year} {2013})\ pp.\ \bibinfo {pages}
  {1007--1010}\BibitemShut {NoStop}%
\bibitem [{\citenamefont {Amrouche}\ \emph {et~al.}(1998)\citenamefont
  {Amrouche}, \citenamefont {Bernardi}, \citenamefont {Dauge},\ and\
  \citenamefont {Girault}}]{AmroucheNonSmooth}%
  \BibitemOpen
  \bibfield  {author} {\bibinfo {author} {\bibfnamefont {C.}~\bibnamefont
  {Amrouche}}, \bibinfo {author} {\bibfnamefont {C.}~\bibnamefont {Bernardi}},
  \bibinfo {author} {\bibfnamefont {M.}~\bibnamefont {Dauge}}, \ and\ \bibinfo
  {author} {\bibfnamefont {V.}~\bibnamefont {Girault}},\ }\href {\doibase
  10.1002/(SICI)1099-1476(199806)21:9<823::AID-MMA976>3.0.CO;2-B} {\bibfield
  {journal} {\bibinfo  {journal} {Mathematical Methods in the Applied
  Sciences}\ }\textbf {\bibinfo {volume} {21}},\ \bibinfo {pages} {823}
  (\bibinfo {year} {1998})}\BibitemShut {NoStop}%
\bibitem [{\citenamefont {{Boyse}}\ and\ \citenamefont
  {{Paulsen}}(1997)}]{BoyseNonConvex}%
  \BibitemOpen
  \bibfield  {author} {\bibinfo {author} {\bibfnamefont {W.~E.}\ \bibnamefont
  {{Boyse}}}\ and\ \bibinfo {author} {\bibfnamefont {K.~D.}\ \bibnamefont
  {{Paulsen}}},\ }\href {\doibase 10.1109/8.650193} {\bibfield  {journal}
  {\bibinfo  {journal} {IEEE Transactions on Antennas and Propagation}\
  }\textbf {\bibinfo {volume} {45}},\ \bibinfo {pages} {1758} (\bibinfo {year}
  {1997})}\BibitemShut {NoStop}%
\bibitem [{\citenamefont {Tisseur}\ and\ \citenamefont
  {Meerbergen}(2001)}]{TisseurQEP}%
  \BibitemOpen
  \bibfield  {author} {\bibinfo {author} {\bibfnamefont {F.}~\bibnamefont
  {Tisseur}}\ and\ \bibinfo {author} {\bibfnamefont {K.}~\bibnamefont
  {Meerbergen}},\ }\href@noop {} {\bibfield  {journal} {\bibinfo  {journal}
  {SIAM Rev.}\ }\textbf {\bibinfo {volume} {43}},\ \bibinfo {pages} {235}
  (\bibinfo {year} {2001})}\BibitemShut {NoStop}%
\bibitem [{\citenamefont {Gavin}\ \emph {et~al.}(2018)\citenamefont {Gavin},
  \citenamefont {Międlar},\ and\ \citenamefont {Polizzi}}]{NLFEAST}%
  \BibitemOpen
  \bibfield  {author} {\bibinfo {author} {\bibfnamefont {B.}~\bibnamefont
  {Gavin}}, \bibinfo {author} {\bibfnamefont {A.}~\bibnamefont {Międlar}}, \
  and\ \bibinfo {author} {\bibfnamefont {E.}~\bibnamefont {Polizzi}},\
  }\href@noop {} {\bibfield  {journal} {\bibinfo  {journal} {Journal of
  Computational Science}\ }\textbf {\bibinfo {volume} {27}},\ \bibinfo {pages}
  {107 } (\bibinfo {year} {2018})}\BibitemShut {NoStop}%
\bibitem [{COM()}]{COMSOL}%
  \BibitemOpen
  \href@noop {} {\emph {\bibinfo {title} {COMSOL Multiphysics Reference
  Manual}}}\BibitemShut {NoStop}%
\bibitem [{\citenamefont {Zienkiewicz}\ and\ \citenamefont {Zhu}(1992)}]{spr}%
  \BibitemOpen
  \bibfield  {author} {\bibinfo {author} {\bibfnamefont {O.}~\bibnamefont
  {Zienkiewicz}}\ and\ \bibinfo {author} {\bibfnamefont {J.}~\bibnamefont
  {Zhu}},\ }\href {\doibase https://doi.org/10.1016/0045-7825(92)90023-D}
  {\bibfield  {journal} {\bibinfo  {journal} {Computer Methods in Applied
  Mechanics and Engineering}\ }\textbf {\bibinfo {volume} {101}},\ \bibinfo
  {pages} {207 } (\bibinfo {year} {1992})}\BibitemShut {NoStop}%
\bibitem [{\citenamefont {Mur}(1994)}]{GMurAdvDis}%
  \BibitemOpen
  \bibfield  {author} {\bibinfo {author} {\bibfnamefont {G.}~\bibnamefont
  {Mur}},\ }\href {\doibase 10.1109/20.312706} {\bibfield  {journal} {\bibinfo
  {journal} {IEEE Transactions on Magnetics}\ }\textbf {\bibinfo {volume}
  {30}},\ \bibinfo {pages} {3552} (\bibinfo {year} {1994})}\BibitemShut
  {NoStop}%
\end{thebibliography}%

\appendix*
\section{Variation of the Field Theory Lagrangian}
\label{sec:VariationAppendix}
The Lagrangian for the electromagnetic four-potential including the gauge fixing term and free sources is given by eq. \ref{eq:CovariantL}. For simplicity we assume constant permeability and permitivitty, $\mu$ and $\epsilon$ in a given mesh element thus neglecting derivatives of these parameters, however future work could consider extending this to anisotropic heterogeneous materials even within a mesh element. We also work with $\alpha = \frac{1}{2\xi}$ as the coefficient of the gauge fixing term to keep the notation clean. Finally we are using the metric signature ($+---$) in the following.
\begin{equation}
\mathcal{L}= -\frac{1}{4 \mu} F^{\nu \beta}F_{\nu \beta} - A_\nu J^\nu - \frac{\alpha}{2} 
\left(\partial_{\nu} A^{\nu}\right)^2 \label{eq:CovariantL}
\end{equation}

Expanding the four-potential in terms of the components, $\mathbf{A} = (\frac{\phi}{c}, \vec{A}$) where $c$ is the speed of light in the medium and, as we are working with time harmonic potentials in the frequency domain, replacing derivatives with respect to time by $i \omega$, we obtain eq.~\ref{eq:Covariant3P1L}.
\begin{multline}
\mathcal{L}=\frac{1}{2} [\epsilon (\nabla \phi + i \omega \vec{A}) \cdot (\nabla \phi^* - i \omega \vec{A}^*) - \frac{1}{\mu} (\nabla \times \vec{A}) \cdot (\nabla \times \vec{A}^*)\\
- \frac{\alpha}{\mu} (\nabla \cdot \vec{A}+i \frac{\omega}{c^2} \phi )(\nabla \cdot \vec{A}^*-i \frac{\omega}{c^2} \phi^* ) - \rho \phi ^*- \rho^* \phi + \vec{A} \cdot \vec{J}^*+ \vec{A}^* \cdot \vec{J}] \label{eq:Covariant3P1L}
\end{multline}

From the point of view of numerical stability and so as to work with parameters having the same dimensions, it is better to normalize $\phi$ by $c_0$, the speed of light in a vacuum, and work with $k_0 = \frac{\omega}{c_0}$ instead of $\omega$.  We thus change to the variable $\tilde{\phi} = \frac{\phi}{c_0}$. Similarly for the space charge density, let us define $\tilde{\rho} = \frac{\rho}{c_0 \epsilon_0}$. Finally, let us define a normalized field $\tilde {\vec{E}} = \frac{\vec{E}}{c_0} = -i k_0 \vec{A} - \nabla \tilde{\phi}$. The units of $\nabla \tilde{\phi}$ and $k_0 \vec{A}$ are now both [V][s][m]$^{-2}$.

This change of variable results in a common factor of $2 \mu_0$ over all terms in eq~\ref{eq:Covariant3P1L} other than the source terms. We divide the entire expression by this factor and subsequently ignore it as in the subsequent analysis we are only interested in setting the variation of this to zero. This gives eq.~\ref{eq:Covariant3P1L2} where $\epsilon_r$ and $\mu_r$ are the relative permitivitty and permeability. The refractive index, $n=\sqrt{\epsilon_r \mu_r}=c_0/c$ now appears in the gauge fixing term as we normalized by $c_0$ rather than $c$ (so as to work with the wavenumber in free space). 
\begin{multline}
\mathcal{L}=\epsilon_r (\nabla \tilde{\phi} + i k_0 \vec{A}) \cdot (\nabla \tilde{\phi}^* - i k_0 \vec{A}^*) - \frac{1}{\mu_r} (\nabla \times \vec{A}) \cdot (\nabla \times \vec{A}^*)\\
- \frac{\alpha}{\mu_r} (\nabla \cdot \vec{A}+i k_0 n^2  \tilde{\phi})(\nabla \cdot \vec{A}^*-i k_0 n^2 \tilde{\phi}^*)\\
- \tilde{ \rho} \tilde{\phi}^* - \tilde{ \rho}^* \tilde{\phi} + \mu_0 \vec{A} \cdot \vec{J}^* +  \mu_0 \vec{A}^* \cdot \vec{J} \label{eq:Covariant3P1L2}
\end{multline}

The full action integral comprising the variational formulation, including the surface impedance boundary integral, is given in eq.~\ref{eq:FullActionInt}. We are interested in taking the variation over the closed volume $\Omega$ with a perfect magnetic boundary on the surface $\Gamma_\mathrm{PM}$ and an impedance boundary on the surface $\Gamma_\mathrm{Z}$. The factor $\gamma=\frac{1}{i \omega \epsilon_0 Z_0}$, the reason for which will be made clear at the end of this appendix.
\begin{multline}
S=\int_\Omega \epsilon_r (\nabla \tilde{\phi} + i k_0 \vec{A}) \cdot (\nabla \tilde{\phi}^* - i k_0 \vec{A}^*) - \frac{1}{\mu_r} (\nabla \times \vec{A}) \cdot (\nabla \times \vec{A}^*)\\
- \frac{\alpha}{\mu_r} (\nabla \cdot \vec{A}+i k_0 n^2  \tilde{\phi})(\nabla \cdot \vec{A}^*-i k_0 n^2 \tilde{\phi}^* )\\
-\tilde{ \rho} \tilde{\phi}^* - \tilde{ \rho}^* \tilde{\phi} +  \mu_0 \vec{A} \cdot \vec{J}^* +  \mu_0 \vec{A}^* \cdot \vec{J} ~\mathrm{d}V+\\
\gamma \int_{\partial \Gamma_\mathrm{Z}}  \left( \hat{n} \times \left(-\nabla \tilde{\phi}- i k_0 \vec{A} \right) \right) \cdot \left(\hat{n} \times \left(-\nabla \tilde{\phi}^*+ i k_0 \vec{A}^*\right) \right)~\mathrm{d}S
\label{eq:FullActionInt}
\end{multline} 

The variation of eq.~\ref{eq:FullActionInt} with respect to $\tilde{\phi}^*$ is taken first:
\begin{multline}
\delta_{\tilde{\phi}^*} S = \int_\Omega \epsilon_r \left(\nabla \tilde{\phi}+ i k_0 \vec{A}\right) \nabla (\delta \tilde{\phi}^*)-\tilde{\rho} \delta \tilde{\phi}^*\\
 +\frac{ i k_0 n^2 \alpha}{\mu_r} \left(\nabla \cdot \vec{A} + i k_0 n^2\tilde{\phi} \right) \delta \tilde{\phi}^*~dV\\
- \gamma \int_{\Gamma_\mathrm{Z}} \left( \hat{n} \times \left(-\nabla \tilde{\phi} - i k_0 \vec{A}\right) \right) \cdot \left(\hat{n} \times \nabla \delta \tilde{\phi}^* \right) ~dS
\end{multline}
\begin{multline}
=  \epsilon_r  \int_\Omega\left(i k_0 (\alpha-1) \nabla \cdot \vec{A} - \alpha n^2 k_0^2  \tilde{\phi} - \nabla^2 \tilde{\phi} - \frac{\tilde{\rho}}{\epsilon_r}\right) \delta \phi^*~dV \\
+\epsilon_r \int_{\partial \Omega}  \left(\nabla \tilde{\phi} + i k_0 \vec{A}\right) \delta \tilde{\phi}^*~dS  \\
+\gamma  \int_{\Gamma_\mathrm{Z}}  \left[(-\nabla \tilde{\phi} - i k_0 \vec{A}) - \left((-\nabla \tilde{\phi} - i k_0 \vec{A})  \cdot \hat{n} \right) \hat{n} \right] \nabla (\delta \tilde{\phi}^*) ~dS
\end{multline}

Where we applied the divergence theorem and the identity $(\vec{A} \times \vec{B}) \cdot (\vec{C} \times \vec{D}) = (\vec{A} \cdot \vec{C})(\vec{B} \cdot \vec{D}) - (\vec{A} \cdot \vec{D})(\vec{B} \cdot \vec{C})$ to the impedance integral. Let us define the projection of $\tilde{\vec{E}}$ on the tangent surface as $\tilde{\vec{E}}_t' = \tilde{\vec{E}}- (\tilde{\vec{E}} \cdot \hat{n}) \hat{n}$ (note that this is orthogonal to $\hat{n} \times \tilde{\vec{E}}$). Next we apply the vector identity, $\vec{B} \cdot \nabla \alpha=\nabla \cdot (\alpha \vec{B}) - \alpha \nabla \cdot \vec{B}$ to the impedance surface integral:
\begin{multline}
\delta_{\tilde{\phi}^*} S =  \epsilon_r  \int_\Omega\left(i k_0 (\alpha-1) \nabla \cdot \vec{A} - \alpha n^2 k_0^2  \tilde{\phi} - \nabla^2 \tilde{\phi} - \frac{\tilde{\rho}}{\epsilon_r}\right) \delta \phi^*~dV \\
+\epsilon_r \int_{\partial \Omega}  \left(\nabla \tilde{\phi} + i k_0 \vec{A}\right) \delta \tilde{\phi}^*~dS  + \gamma \int_{\Gamma_\mathrm{Z}} \nabla \cdot (\tilde{\vec{E}}_t' \delta \tilde{\phi}^*) - \nabla \cdot (\tilde{\vec{E}}_t') \delta \tilde{\phi}^* ~dS
\end{multline}

We can then apply the divergence theorem on a surface to convert the first term in the impedance boundary to a line integral which vanishes over a closed surface. The corresponding set of equations enforced through the variation with respect to $\phi^*$ is then given by eqs.~\ref{eq:dSphi1} - ~\ref{eq:dSphi2} where we have given the final system of equations in terms of the unnormalized potential and fields. There is also a corresponding set for the complex conjugate terms, obtained through the variation with respect to $\tilde{\phi}$.
\begin{align}
\alpha n^2 k_0^2 \phi+ \nabla^2 \phi - i k_0 (\alpha-1) \nabla \cdot \vec{A} = -\frac{\rho}{\epsilon} & \hspace{5pt} \mathrm{in}~\Omega \label{eq:dSphi1} \\
\left(-\nabla \phi - i k_0 \vec{A}\right) \cdot \hat{n} = \vec{E}_n = 0 & \hspace{5pt} \mathrm{on}~\Gamma_\mathrm{PM}  \\
\epsilon_r E_n + \gamma \nabla \cdot \vec{E}_t'= 0 & \hspace{5pt} \mathrm{on}~\Gamma_\mathrm{Z} \label{eq:dSphi2}
\end{align}

We shall reserve the discussion of these terms until after the variation with respect to $\vec{A}^*$, which we now take.
\begin{multline}
\delta_{\vec{A}^*} S= \int_{\Omega} -i k_0 \epsilon_r (\nabla \tilde{\phi} + i k_0 \vec{A}) \vec{\delta A}^* + \mu_0 \vec{J} \cdot  \vec{\delta A}^* \\
- \frac{\alpha}{\mu_r} (\nabla \cdot \vec{A}+i k_0 n^2 \tilde{\phi}) \nabla \cdot \vec{\delta A}^*- \frac{1}{\mu_r} (\nabla \times \vec{A}) \cdot (\nabla \times \vec{\delta A}^*) ~dV\\
+ i k_0 \gamma \int_{\Gamma_\mathrm{Z}} [\hat{n} \times (-\nabla \tilde{\phi} - i k_0 \vec{A})] \cdot (\hat{n} \times \vec{\delta A}^*) ~dS
\end{multline}
\begin{multline}
= \int_{\Omega} [ -i k_0 \epsilon_r \nabla \tilde{\phi} + \epsilon_r k_0^2 \vec{A} + \mu_0 \vec{J}+ \frac{\alpha}{\mu_r} \nabla (\nabla \cdot \vec{A}+i k_0 n^2 \tilde{\phi})\\
-\frac{1}{\mu_r} \nabla \times (\nabla \times \vec{A}) ] \cdot \vec{\delta A}^*~dV- \frac{1}{\mu_r}\int_{\partial \Omega} \left(\vec{\delta A}^* \times (\nabla \times \vec{A})\right) \cdot \hat{n}~dS\\
+ i k_0 \gamma \int_{\Gamma_\mathrm{Z}} [(-\nabla \tilde{\phi} - i k_0 \vec{A}) - \left((-\nabla \tilde{\phi} - i k_0 \vec{A})  \cdot \hat{n} \right)~ \hat{n}] \cdot \vec{\delta A}^* ~dS\\
- \frac{\alpha}{\mu_r} \int_{\partial \Omega} (\nabla \cdot \vec{A}+i k_0 n^2 \tilde{\phi} ) \vec{\delta A}^* \cdot \hat{n}~dS
\end{multline}
\begin{multline}
= \int_{\Omega} [i k_0 n^2 (\alpha -1)\nabla \tilde{\phi} + \epsilon_r k_0^2 \vec{A}\\
+\frac{1}{\mu_r} (\alpha -1)  \nabla (\nabla \cdot \vec{A} ) + \frac{1}{\mu_r} \nabla^2 \vec{A}] \cdot \vec{\delta A}^*~dV \\
+\frac{1}{\mu_r}\int_{\partial \Omega} ( \hat{n} \times (\nabla \times \vec{A}) - \alpha (\nabla \cdot \vec{A}+i k_0 \tilde{\phi} ) \hat{n} ) \cdot \vec{\delta A}^* ~dS+\\
 i k_0 \gamma \int_{\Gamma_\mathrm{Z}} \vec{E}_t' \cdot \vec{\delta A}^* ~dS
\end{multline}

The total set of equations that are satisfied when $S$ is minimized (including eqs.~\ref{eq:dSphi1} - ~\ref{eq:dSphi2} obtained through the variation with respect to $\tilde{\phi}^*$) are as follows. Once again, we have substituted the original expressions for $\phi$, $\vec{E}$ and $\rho$ into these equations.
\begin{align*}
\alpha n^2 k_0^2 \phi+ \nabla^2 \phi - i k_0 (\alpha-1) \nabla \cdot \vec{A} = -\frac{\rho}{\epsilon} & ~\mathrm{in}~\Omega \\
n^2 k_0^2 \vec{A} +\nabla^2 \vec{A} + (\alpha -1) \nabla (\nabla \cdot \vec{A} + i \frac{\omega}{c^2} \phi) = -\mu \vec{J} &  ~\mathrm{in}~\Omega\\
\left(-\nabla \phi - i c_0 k_0 \vec{A}\right) \cdot \hat{n} = \vec{E}_n = 0 &  ~\mathrm{on}~\Gamma_\mathrm{PM}  \\
\hat{n} \times (\nabla \times \vec{A}) - \alpha \left(\nabla \cdot \vec{A}+i \frac{\omega}{c^2} \phi \right) \hat{n}=0 & ~\mathrm{on}~\Gamma_\mathrm{PM}\\
\epsilon_r E_n + \gamma \nabla \cdot \vec{E}_t' = 0 &~\mathrm{on}~\Gamma_\mathrm{Z}  \\
\hat{n} \times (\nabla \times \vec{A}) - \alpha(\nabla \cdot \vec{A}+i \frac{\omega}{c^2} \phi ) \hat{n} +  \frac{i k_0 \mu_r \gamma }{c_0}\vec{E}_t' =0 &  ~\mathrm{on}~\Gamma_\mathrm{Z}
\end{align*}

Once again, there are another set of equations consisting of the complex conjugate terms, corresponding to the variation with respect to $\tilde \phi$ and $\vec{A}$. When the Lorenz gauge is enforced, $\nabla \cdot \vec{A} + i \frac{\omega}{c^2} \phi =0$ and the boundary conditions imposed by the natural boundary condition are those corresponding to a perfect magnetic conductor.
\begin{align}
\vec{E}_n = 0\\
\hat{n} \times (\nabla \times \vec{A}) = \hat{n} \times \vec{B} = \vec{B}_t = 0
\end{align}

Meanwhile, on the surface with the impedance surface integral added, $\Gamma_\mathrm{Z}$, the boundary condition imposed by the variation with $\vec{A}$ is as follows:
\begin{align}
\vec{E}_t' =  \vec{E}- (\vec{E} \cdot \hat{n}) \hat{n} &= - \frac{c_0}{i k_0 \gamma \mu_r} \vec{B}_t  &= - \frac{c_0 \mu_0}{i k_0 \gamma} \vec{H}_t  = Z \vec{H}_t \hspace{25pt}
\end{align}

From which we determine the relationship between our original $\gamma$ and the complex impedance on the boundary, $Z$:
\begin{equation}
\gamma= -\frac{c_0 \mu_0}{i k_0 Z} = -\frac{1}{i \epsilon \omega Z} 
\end{equation}

\end{document}